\documentclass[useAMS,usenatbib]{mn2e} 
\usepackage{graphicx,lscape}
\usepackage{lineno} 

\def \gx {GX~339--4}

\title[GX 339-4]{The evolution of the high-energy cut-off in the X-ray spectrum of GX 339-4 across a hard-to-soft transition}
\author[S. Motta et al.]{
S. Motta$^{1,2}$, T. Belloni$^{1}$, J. Homan$^{3}$,\\\\
$^{1}$INAF-Osservatorio Astronomico di Brera, Via E. Bianchi 46, I-23807 Merate (LC), Italy\\
$^{2}$Universit\`a dell'Insubria, Via Valleggio 11, I-22100 Como, Italy \\
$^{3}$Center for Space Research, Massachusetts Institute of Technology, 77 Massachusetts Avenue, Cambridge, MA 02139-4307, USA\\
}

\begin{document}

\date{Accepted 2009 August 17. Received 2009 August 17; in original form 2009 June 05
}

\pagerange{\pageref{firstpage}--\pageref{lastpage}} \pubyear{0000}

\maketitle

\label{firstpage}

\begin{abstract}

We report on X-ray observations of the black-hole candidate GX
339-4 during its 2006/2007 outburst. The hardness-intensity diagram of all RXTE/PCA data
combined shows a q-shaped track similar to that observed in
previous outbursts. The evolution through the HID suggests that in the early phase of the outburst the source underwent a sequence of state transitions, from the hard to the soft state, which is supported by our timing analysis. Broadband (4-200 keV) spectra, fitted with an exponentially cutoff powerlaw, show that the hard spectral component steepens during the transition from the hard to the soft state. The high-energy cutoff decreased monotonically from 120 to 60 keV during the brightening of the hard state, but increased again to 100 keV during the softening in the hard intermediate state. In the short-lived soft intermediate state the cutoff energy was $\sim$ 130 keV, but was no longer detected in the soft state. This is one of the first times that the high-energy cut-off has been followed in such detail across several state transitions. We find that in comparison to several other spectral parameters, the cut-off energy changes more rapidly, just like the timing properties. The observed behaviour of the high energy cutoff of GX 339-4 is also similar to that observed with RXTE-INTEGRAL-Swift during the 2005 outburst of GRO J1655-40.
These results constitute a valuable reference to be considered when testing theoretical models for the production
of the hard component in these systems.

\end{abstract}


\section{Introduction}

The spectral evolution of black hole X-Ray transients (BHTs) has recently been described in terms of
patterns in an X-ray hardness-intensity diagram (HID) (see Homan \&
Belloni 2005; Belloni et al. 2005; Belloni 2005, Gierlinski \& Newton 2006, Remillard \& McClintok 2006, Belloni 2009, Fender, Homan, Belloni 2009). 
Different states are found to correspond to different branches/areas of a q-like
 pattern that shows up in a log-log representation. Four main states are identified within this
framework: Low Hard State (LHS), Hard Intermediate State (HIMS), Soft Intermediate State (SIMS), High Soft State (HSS). 
Twostates correspond to the original states
discovered in the 1970s: the Low/Hard State (LHS), observed
usually at the beginning and at the end of an outburst, showing a spectrum dominated by an hard component with sometimes a thermal disc component (very faint) and the  High/Soft State, usually observed in the central intervals of an outburst, that shows an energy spectrum dominated by a thermal disc component, with the presence of an additional weak and steep power-law component.

In between these two well-established states, the situation is rather complex and
has led to a number of different classifications. Homan \& Belloni
(2005) identify two additional states, clearly defined by
spectral/timing transitions. In the evolution of a transient,
after the LHS comes a transition to the Hard Intermediate State: the energy spectrum softens as the combined result of a
steepening of the power-law component and the appearance of a
thermal disc component. At the same time, the characteristic
frequencies in the power spectrum increase and the total
fractional rms decreases. The transition to
the Soft-Intermediate State can be very fast (sometimes over a few
seconds, see Nespoli et al. 2003) and is marked by the
disappearance of some particular features in the power density spectrum and by the appearance of some others. 

Together with the association of the transition to the SIMS with the ejection
of fast relativistic jets, this has led to the identification of a
\emph{jet line} in the HID, separating HIMS and SIMS (Fender,
Belloni \& Gallo 2004). The jet line can be crossed more than once
during an outburst (as in the case of XTE J1859+226:  Casella et
al. 2004; Brocksopp et al. 2002). Notice that recently the comparative 
study of different systems has shown that the jet ejection and HIMS/SIMS state transitions 
are not exactly simultaneous (Fender, Homan, Belloni 2009). For a more detailed 
state classification see Belloni (2009). 
McClintock et al. (2009 and references therein) use another state classification  based more on spectral properties than on timing properties, unlike the classification presented by Belloni \& Homan.They define three different states on the basis of precise boundaries of a number of parameters such as integrated fractional rms and the presence of QPO in the Power Density Spectrum, power law photon index and disk fraction in the energy spectra. For a comparison between the two classifications, see Belloni (2009).



While the physical nature of the soft component  in the X-Ray spectra of BHTs is commonly
associated with an optically thick accretion disc, there is no
consensus as to the origin of the hard spectral component. Nevertheless there are various suggestions regarding this: the hard spectral component could be due to the presence of different component, such as a hot corona, the very inner part of the accretion flow, the formation/ejection of relativistic jets. The hard component is usually interpreted as the result of thermal Comptonization or of a combination of thermal/non-thermal Comptonization involving the hot electrons of the corona and the soft photons originating in the accretion disk.
 When fitted with a power-law, the slope of the hard component is typically found to be 
$\sim$1.6 for the LHS, 1.6-2.5 in the SIMS/HIMS and 2.5-4 for the
HSS (Belloni et al. 2006). 
It has been known for a long time (Sunyaev \& Trumper 1979) that the LHS spectrum shows a cutoff around $\sim$ 100 keV. A comparative measurement of a number of systems with CGRO/OSSE has been presented by Grove et al. (1998). Here the energy spectra could be clearly divided into two classes: the ones with a strong soft thermal component and no evidence of a high-energy cutoff until $\sim$ 1 MeV, and those with no soft component and a $\sim$ 100 keV cutoff. Zdziarski et al. (2001) and Rodr\'iguez et al. (2004) measured the high-energy spectrum of GRS 1915+105 and found no direct evidence of a high-energy cutoff, but spectra which appeared to contain two components. At energies of 50-150 keV, the hard spectral component often shows a cut off (Belloni et al. 2006, Joinet et al. 2008, Miyakawa et al. 2008, Del Santo et al. 2008), which can provide additional information about the properties and origin of the hard spectral component. This cut off is thought to be related to the temperature of the thermal comptonizing electrons located in an optically thin corona close to the black hole, responsible for the comptonization of the soft photons emitted by the accretion disc.  
More recently Miyakawa et al. (2008)  performed an analysis on Rossi XTE Observation of GX 339-4 with the aim to investigate the radiation mechanism in the hard state of the source. They observed a high-energy cutoff ranging from 40 to over 200 keV. Joinet et al. (2008) presented the analysis of the high energy emission of GRO J1655-40 at the beginning of the 2005 Outburst. Their high-energy data allowed them to detect the presence of a high-energy cutoff and to study its evolution during the outburst rise. They observed a cutoff decreasing from above 200 keV down to $\sim$ 100 keV. Following that it either increases significantly or vanishes completely. Caballero-Garcia et al. (2009) also studied GRO J1655-40 during the 2005 outburst but claimed that no cutoff was required for their INTEGRAL dataset.


The X-ray spectra of BHTs  also include additional components which are 
important in terms of the physics of accretion onto black holes, such as
emission and absorption line features  (see e.g. Reynolds \& Nowak 2003, Miller
et. al 2002, Miller et al., 2004a, Miller et. al 2006, Neilsen \& Lee 2009) and Compton reflection humps
(see e.g. George \& Fabian 1991, Zdziarski et al. 2001, Frontera
et al. 2001). 

\subsection{GX 339-4}

GX 339--4 was one of the first two BHTs for which a complete set
of transitions was observed and studied (see Miyamoto et al.
1991; Belloni et al. 1997; M\'endez \& van der Klis 1997), and is known to spend long periods in
outbursts. 
A detailed study of the evolution of the hard spectral component in GX 339-4 at energies above 3 keV was performed during its 2004 outburst which began in February of that year.  To get
broad-band coverage during the expected HIMS-SIMS spectral transition,
simultaneous RXTE and INTEGRAL observations were made. Belloni et
al. (2006) combined data from PCA, HEXTE and IBIS, and obtained good
quality broad-band (3-200 keV) energy spectra before and soon-after the
transition. These spectra indicated steepening of the hard, high-energy
component. Also, the  high-energy cut-off which was present at $\sim$70
keV before the transition was not detected later. Therefore, although
spectral parameters at lower energies do not change abruptly during the
transition, the energy of the cut-off increases or disappears  rapidly (within 10 hours). The power spectra before and after the
transition showed significant differences (see Belloni et al. 2005;
Belloni 2008): from strong band-limited noise and type-C QPO to much
weaker noise and type-B QPO (for a description of the properties of
different types of QPO, see Casella, Belloni \& Stella 2005).


Del Santo et al. (2008) report on X-ray and soft $\gamma$-ray observations of the black-hole
candidate during an outburst in 2006/2007, performed with the RXTE and
INTEGRAL satellites. The evolution in the HID of all RXTE/PCA 
data suggests that a transition from hard-intermediate state to soft-intermediate
state occurred, simultaneously with  INTEGRAL observations performed in March.
The transition was confirmed by the timing analysis which
revealed that a weak type-A quasi-periodic oscillation (QPO) replaced a strong type-C QPO.
At the same time, spectral analysis revealed that the flux of the high-energy component showed 
a significant decrease. However, Del Santo et al. observed a delay of roughly one day 
between variations of the spectral parameters of the high-energy component
and changes in the flux and timing properties.

The aim of this work is to use RXTE data collected during the 2006/2007
outburst to study the broad-band spectral evolution of GX 339-4
during a full hard-to-soft state transitions and in particular the
behaviour of the high-energy cut off. 
Here we study the 
spectral evolution of GX 339-4 over a longer period
of time, covering almost the entire LHS to HSS transition
(from 27 December 2006 to 18 April 2007).  Thanks to the unprecedented data coverage
of the main phases of the source's evolution we were able for the
first time to follow in a detailed way the spectral evolution of the
source and in particular of the cut-off energy component of the spectra over
the LH, HIMS, SIMS and the first part of the HSS state of the
sources.

\section{Observations and data analysis}\label{sec:observations} 

In  2006 November, X-ray activity of GX 339-4 was detected with the
Rossi X-ray Timing Explorer (RXTE; Swank et al. (2006)). The
source had an almost constant flux until the end of December 2006,
when the hard (15-50 keV) X-ray flux increased by a large amount.
It reached its brightest level since 2004 November, as detected by
SWIFT/BAT (Krimm et al. 2006).  In order to follow the new
outburst of GX 339-4 at high energies, an RXTE ToO campaign was
carried out. 
As expected in BHTs (Homan $\&$ Belloni 2005),
at the beginning of the outburst GX 339-4 was in the LHS. Results of X-ray and soft $\gamma - ray$
observations of GX 339-4 during its 2006/2007 outburst, performed with
RXTE and INTEGRAL satellites have already been reported in Del
Santo et al (2008). Additional results can be found in Caballero-Garcia
et al (2009), who reported on simultaneous XMM-Newton
and INTEGRAL observations.

Starting from December 27, 2006 (MJD 54096) a  total of 220
RXTE pointings were performed  over a period of about 10 months, covering the
full outburst of the source. We
report here the color analysis of all observations (MJD 54096 to 54388, see Figs. \ref{fig:licu} and \ref{fig:HID}, top panel) and spectral analysis of 83  observations (MJD 54134 to 54208) covering the
transition from the LHS to HSS. 

We extracted energy spectra from the PCA and HEXTE instruments (background and deadtime
corrected) for each observation using the standard RXTE software
within HEASOFT V. 6.4, following the standard extraction
procedures. For our spectral analysis, only Proportional Counter Unit  2 from the PCA and
Cluster B from HEXTE were used. A systematic error of $0.6\%$
was added to the PCA spectra to account for residual
uncertainties in the instrument calibration. We accumulated
background corrected PCU2 rates in the channel bands A = 4 - 45
(3.2 - 18.3 keV), B = 4 - 10 (3.2 - 5.4 keV) and C = 11 - 20 (5.7
- 9.0 keV). A is the total rate, while the hardness was defined
as H = C/B (see Homan $\&$ Belloni 2005). PCA+HEXTE spectra were fitted with XSPEC V. 11 in the energy range 3 - 20 keV and 20-200 keV respectively. See Tab. \ref{tab:colori} for the counts and hardness ratio values.

For our timing analysis, we used custom software under
IDL. For each observation we produced power density spectra (PDS) from stretches
128 seconds long using two separate energy bands: PCA channel band
0-249 (corresponding to $\sim$ 2-60 keV)  for the main power spectrum,
and PCA channel band 18-249  ($\sim$ 7.7-60 keV) in order to look for
high-frequency oscillations, which are usually more prominent in
this high-energy band (see Homan et al. 2002). We averaged the power spectra and subtracted the
contribution due to Poissonian noise (see Zhang et al. 1995) in
order to produce two Power Density Spectra (PDS) for each
observation: one for the whole energy band and one for the high
one. The power spectra were normalized according to Leahy et al.
(1983) and converted to squared fractional rms (Belloni $\&$
Hasinger 1990). See Tab. \ref{tab:colori} for the rms values.

\section{Results}

\begin{figure}
\begin{center}
\includegraphics[height= 4.41cm,width= 8.5cm]{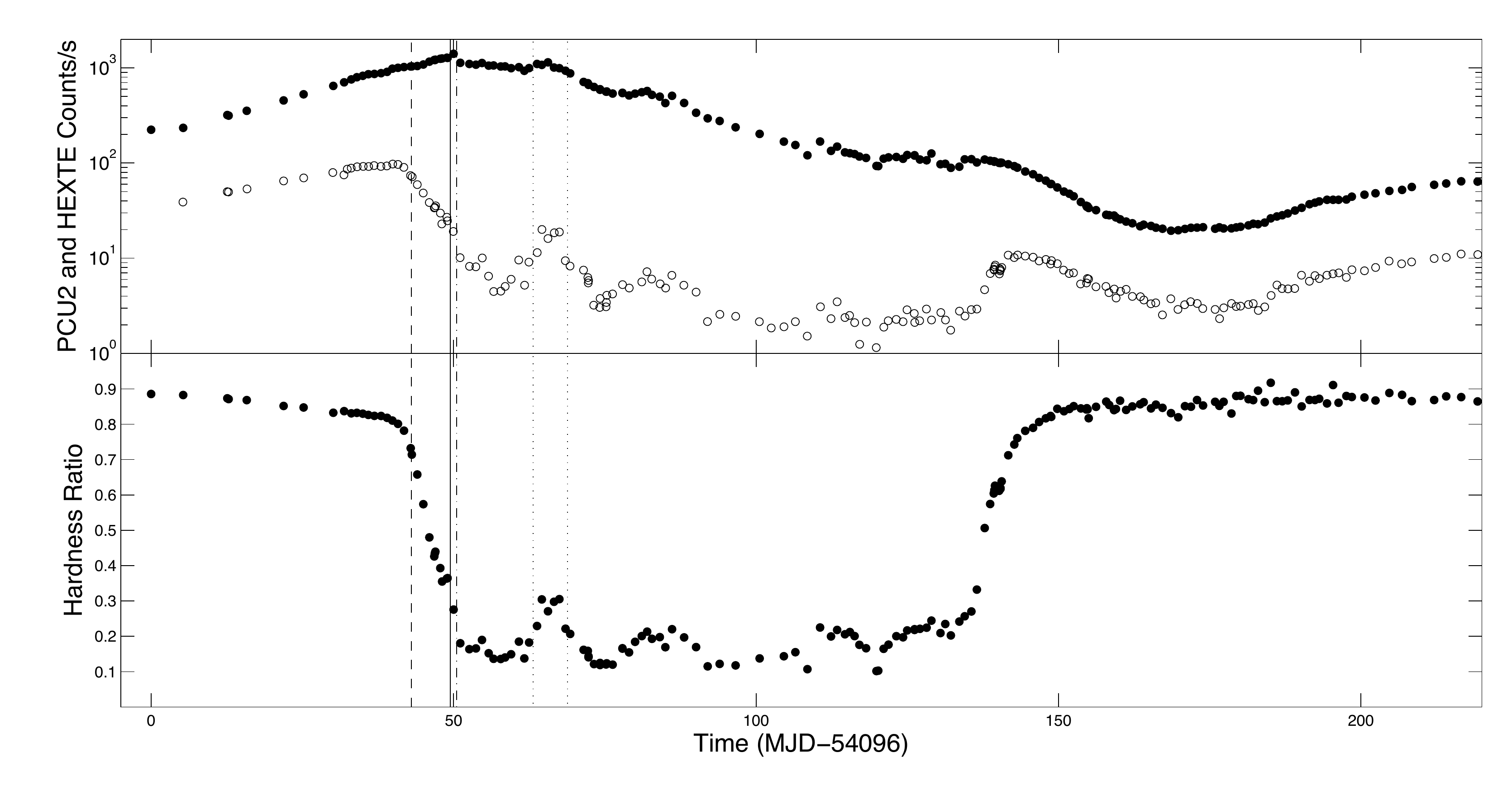}
\caption{Top panel: RXTE/PCA (filled circles) and RXTE/HEXTE (empty circles) light curve of GX 339-4 during the 2006/2007 outburst. The energy range is  3.3 - 21.0 keV for PCA and  19.0 - 200.0 keV for HEXTE. The  vertical lines separate the four canonical States (see Sec. \ref{sec:timing} and \ref{sec:discussion}). Bottom panel: Time evolution of the hardness ratio. The dashed line separates the LHS from the HIMS, the solid line marks the passage from the HIMS to the SIMS and the dot-dashed line separates the SIMS from the HSS. Notice that the source crosses the HIMS-SIMS transition line several times; we marked only the first transition from the HIMS to the SIMS and from the SIMS to the HSS. The two dotted lines indicate the time interval during which the source undergoes several transitions from and to the SIMS (See Sec. \ref{sec:discussion}).}\label{fig:licu}
\end{center}
\end{figure}

\begin{figure}
\begin{center}
\includegraphics[height= 6.41cm,width= 8.5cm]{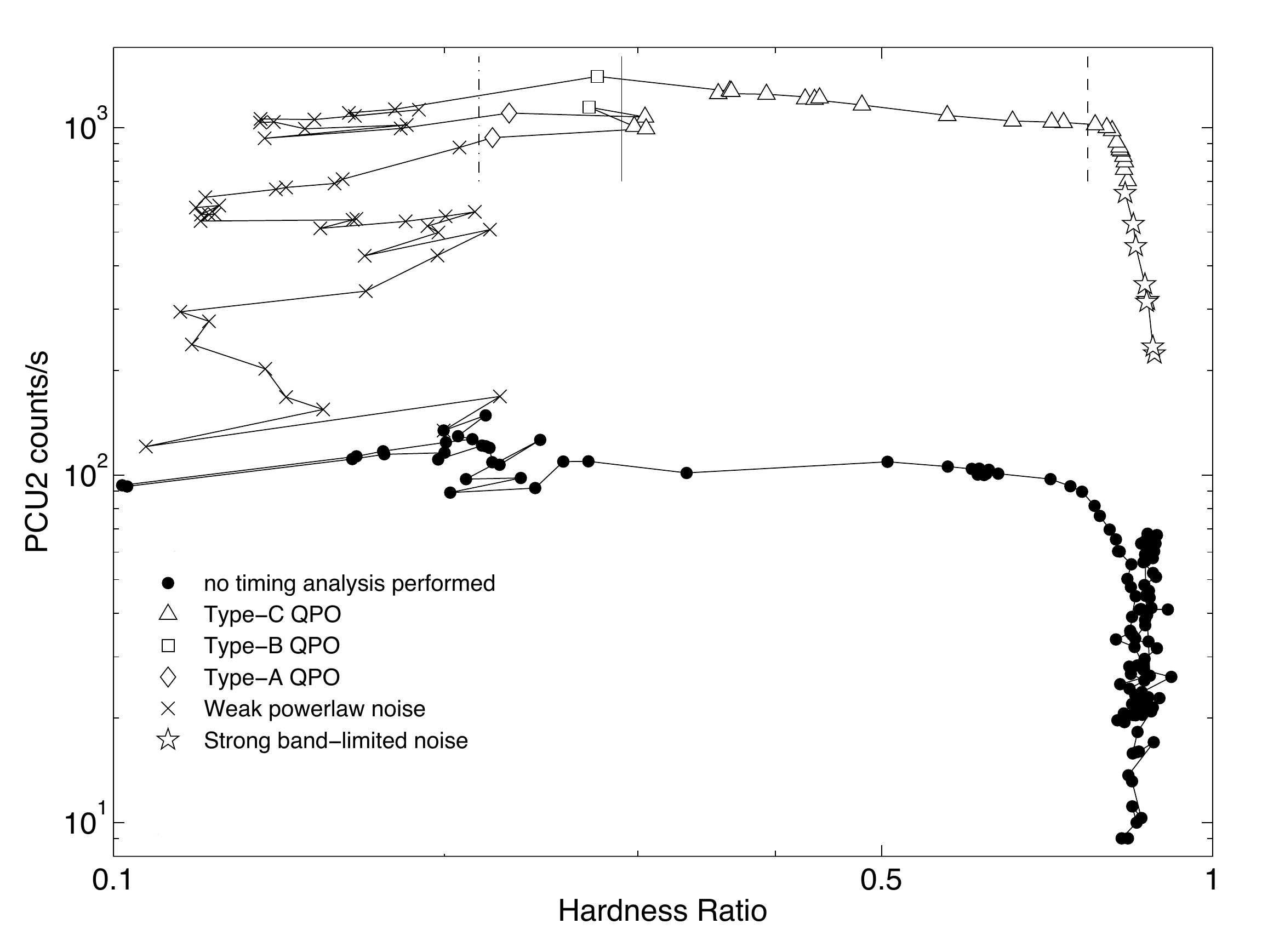}
\caption{Hardness-Intensity diagram from RXTE/PCA data for the complete 2006/2007 outburst which starts from the middle right and proceeds in a counter-clockwise direction. Different symbols indicate different timing properties: type-A QPOs (diamonds), type-B QPOs (squares), type-C QPOs (triangles), strong band-limited noise components in the Power Density Spectrum (stars), weak powerlaw noise in the Power Density Spectrum (crosses). The black dots indicate observations for which we did not perform timing analysis. The vertical lines mark the
transitions as in Fig. \ref{fig:licu}. In this plot the SIMS-HSS transition line is the same for both the main and the second SIMS-to-HSS transition (see Sec. \ref{sec:discussion}). }\label{fig:HID}
\end{center}
\end{figure}

In this section, we describe the general evolution of the
outburst. In Fig. \ref{fig:licu}, we show the full light curve of
the outburst (top panel) and the evolution of the hardness ratio (bottom panel). The Hardness Intensity diagram is shown in Fig. \ref{fig:HID}. 

In Tab. \ref{tab:colori} we list the background corrected PCU2 count rate and the hardness. 
As one can see from Fig. \ref{fig:licu} and Fig. \ref{fig:HID},  the source evolution during the outburst is very similar to that
of the 2002/2003 and 2004 outbursts (see also Belloni et al. 2005): a monotonic
increase in count rate at a rather high color, a horizontal branch with the source softening at a nearly
constant count rate, softening with a
transition to the SIMS, and further observations at very low
hardness. Finally, at count rates lower than the initial LHS-HSS transition, the transition from the HSS back to the LHS takes place.
A noticeable difference between
this outburst and the 2004 outburst is the count rate level of the
top horizontal branch, which is a factor of $\sim$
3.5 higher in the 2006/2007 outburst, similar to the count rate level of the
2002/2003 outburst.

In this work we concentrate on the first part of the outburst, covering the LHS and the complete LHS to HSS transition, from observation \#1 to observation \#83. We will refer to Tab. \ref{tab:colori} and Tab. \ref{tab:spettrali} and to the observation numbers reported therein.

\begin{figure}
\begin{center}
\includegraphics[height= 5.77cm,width= 9cm]{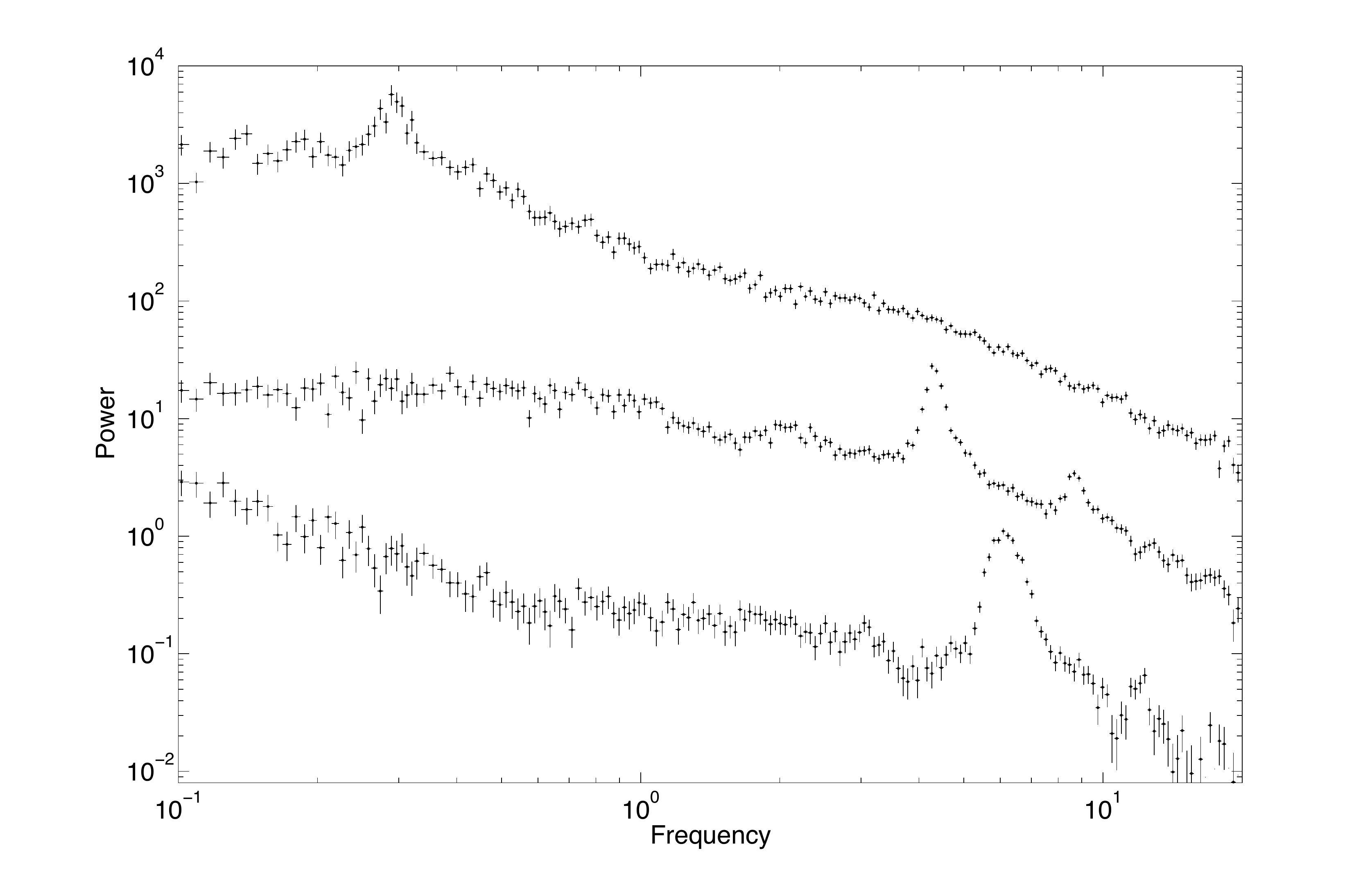}
\caption{Power density spectra for PCA belonging to two different states. Top curve: Obs. $\#15$, belonging to the LHS (right vertical branch in the HID, Fig. \ref{fig:HID}). Middle curve: Obs. \#25, belonging to the HIMS (top horizontal branch in the HID). Type-C QPOs are evident in both PDS. Bottom curve: obs. $\#47$, belonging to the SIMS, middle part of the top horizontal branch of the HID in Fig. \ref{fig:HID}. A strong type B QPO is evident in the PDS. We observed 2 Type B QPOs and 2 Type A QPOs during the SIMS. The top curve is multiplied by a factor of 10 and the bottom curve is scaled down by a factor of 10 for clarity.}\label{fig:power_Spectra}
\end{center}
\end{figure}

\subsection{Timing Analysis}\label{sec:timing}

Since we usually describe the evolution of a source in terms of spectral states, defined on the basis of spectral and timing properties, first of all we have to classify all observations following stated criteria (see Homan \&
Belloni 2005; Belloni et al. 2005; Belloni 2005, Belloni 2009, Fender, Homan, Belloni 2009). For this reason we need the HID and the timing informations in order to identify the branches we see in the HID in terms of canonical BHTs canonical states, which will serve as a framework for the spectral analysis (see Sec. \ref{sec: spectral_analysis}).
Beginning form the HID and examining the power spectra, the presence of state transitions becomes more clear. 

\begin{itemize}

\item Observations from $\#1$ to $\#19$ show a high level of aperiodic variability in the form of strong band-limited noise components  (see Fig. \ref{fig:power_Spectra},  top curve), with total integrated fractional rms in the range  $26 - 47\%$, positively correlated with hardness. The PDS can be decomposed in a number of Lorentzian components, one of which can take the form of a type-C QPO peak (see table \ref{tab:timing_tab}). 
All the observations \#1 to \#19 correspond to the right branch in the top panel of Fig.\ref{fig:HID}. The total fractional rms decreases as the source softens and brightens. All observations in which we found type-C QPOs are marked with triangles in Fig. \ref{fig:HID}, while the other observations are marked with stars.

\item Observations $\#20$ to $\#31$ and observations \#46, \#48, \#49 show fast aperiodic variability with a band-limited noise and strong type-C QPOs (see Fig. \ref{fig:power_Spectra}, middle curve).  The PDS can be decomposed in the same Lorentzian components as in the preceding observations. The total fractional rms is lower than in the LHS (14-23$\%$) and decreases as the source spectrum softens. The observations  correspond to the first part of the horizontal branch in Fig. \ref{fig:HID}, where the largest color variations are observed. Notice that the boundary between this group and the former is somewhat arbitrary, as the evolution in parameters is rather continuous.

\item Observation $\#32$ and $\#47$ correspond to a much weaker variability, in the form of a weak ($\sim 11\%$ fractional rms) powerlaw component. A type-B QPO is prominent in the PDS (see Fig. \ref{fig:power_Spectra}, bottom curve). As observed in the 2004 outburst (see Belloni et al. 2005; Belloni 2008) we see significant differences in the power spectra, with respect to the previous power spectra. Strong band limited noise and a Type-C QPO give way to a much weaker noise and Type-B QPO. Observations \#45 and \#50 show a Type-A QPO, weaker than the type-B QPO detected and with a total fractional rms of  $\sim 5\%$ and $\sim 6\%$ respectively. These two observations were also softer than the two showing a Type-B QPO.  All the observations presenting a type-A or B QPO are marked respectively with diamonds and squares in Fig. \ref{fig:HID}. 

\item Observations from \#33 to \#44 and \#51 to \#83 show weak powerlaw noise with a rms of a few \%. They correspond to the softest observations in Fig. \ref{fig:HID} (top panel). All these observations are marked with crosses in the HID in Fig. \ref{fig:HID}.

\end{itemize}

These results are summarized in Tab. \ref{tab:timing_tab}.

\begin{table*}
\renewcommand{\arraystretch}{1.3}
\begin{center}
\begin{tabular}{|c|c|c|c|}
\hline
\#obs ID 			&    	Noise Type	&	QPO type	&	RMS (in \%)		\\
\hline
\hline

\#1 to \#8					&      strong band limited noise			&		-	&	26-47\%		\\
\#9 to \#19				&      strong band limited noise			&		C	&	26-47\%		\\
\#20 to \#31; \#46,\#48,\#49 	&	 strong band limited noise			&		C	&	14-23\%		\\
\#32,\#47					&	weak powerlaw component		&		A	&	11\%		\\
\#45,\#50					&	weak powerlaw component		&		B	&	5-6\%		\\
\#33 to \#44; \#51 to \#83		&	weak powerlaw component		&		-	&	1-2\%		\\

\hline
\end{tabular}
\caption{Timing properties seen in the PDS of each observation. }\label{tab:timing_tab}
\end{center}
\end{table*}

From the PDS described above, we can identify the four groups of observations as belonging to the LHS (Observations from $\#1$ to $\#19$, right vertical branch of the HID in Fig. \ref{fig:HID}), the HIMS (Observations from $\#20$ to $\#31$ and observations \#46, \#48, \#49, on the right part of the top horizontal branch of the HID in Fig. \ref{fig:HID}), the SIMS (observations $\#32, \#45, \#47, \#50$, marked with squares in the middle of the HID in Fig. \ref{fig:HID}) and HSS (Observations from \#33 to \#44 and observations \#51 to \#83, left vertical branch of the HID in Fig. \ref{fig:HID}). Corresponding transition lines are shown in Fig. \ref{fig:HID} and in Fig. \ref{fig:licu}.
We examined all power spectra, both at low and high energies, for high-frequency features, but found no significant excesses.

In order to compare our results with the classification of McClintock et al. (2009), we follow their updated recipe. They present three states: Hard, Thermal Dominant (TD) and Steep Power Law (SPL). These three states do not fill the complete parameter space: observations which do not qualify are classified as ``intermediate".
For the 83 observations we analyzed in detail, we calculated the disk fraction of the total 2-20 keV unabsorbed flux, the fractional rms integrated over the range 0.1-10 Hz, the QPO amplitude if any QPO is seen and we used the values of photon index for the power-law used in the fits of our spectra (see Tab. \ref{tab:spettrali}).  The criteria used for the McClintock \& Remillard classification are summarized in Tab. 1 in McClintock et al. (2009). In Tab. 1, in addition to our state, we also report theirs. A comparison between the two schemes as applied to these observations can be seen in Tab. \ref{tab:remillard}. 
We see that 42\% of the observations fall into the intermediate state. The others show a general (expected) trend: all the LHS are Hard, most of the HSS are Thermal-Dominant and the SIMS are steep-power law. However, the association is not one to one.  
In conclusion we can say that  for our data the classification we used and McClintock and Remillard classification are quite similar, but for the latter more than $40\%$ of the observations remain unclassified. 

\subsection{Spectral Analysis}\label{sec: spectral_analysis}

PCA (4-20 keV) and HEXTE (20-200 keV) spectra were combined for our  broad-band spectral analysis. For fitting the spectra, we 
used XSPEC V.11.2.3.

\begin{figure}
\begin{center}
\includegraphics[height= 6 cm,width= 8.5 cm]{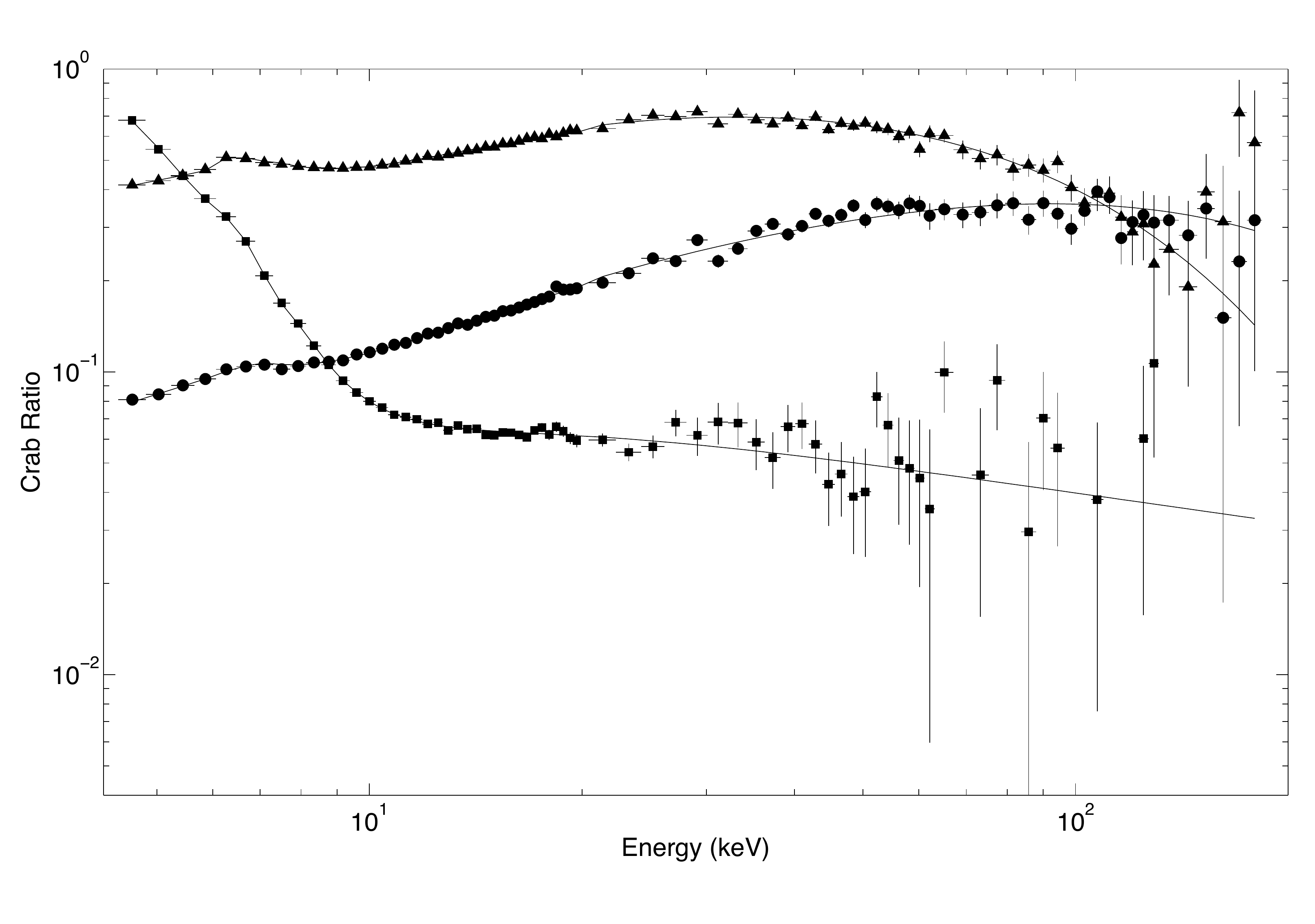}
\caption{PHA ratio of GX 339-4 to Crab for three selected observations. We reported observation \#1 (circles), observation \# 19 (triangles) and observation \#34 (squares).  The high energy cutoff depends on the X-Ray luminosity and decreases with increasing luminosity. }\label{fig:crab_ratio}
\end{center}
\end{figure}

Following Miyakawa et al. (2008) in Fig. \ref{fig:crab_ratio} we first plot the ratio of the three representative spectra to a Crab-like spectrum. We selected Obs. \#1, Obs. \#19 (where we observed respectively the highest and the lowest high-energy cutoff during the LHS)  and Obs. \#34 (where there is not a detectable cutoff). The Crab spectrum was simulated using a simple powerlaw with photon index 2.1 and normalization 10 using {\tt xspec}.  As can be seen from the figure, there is a cutoff at high energies that changes from $\sim 120$ keV (Obs. \# 1) to $\sim 60$ keV (Obs. 19) with increasing flux. 

\begin{figure}
\centering
 \includegraphics[height= 7.4 cm,width= 8.50cm]{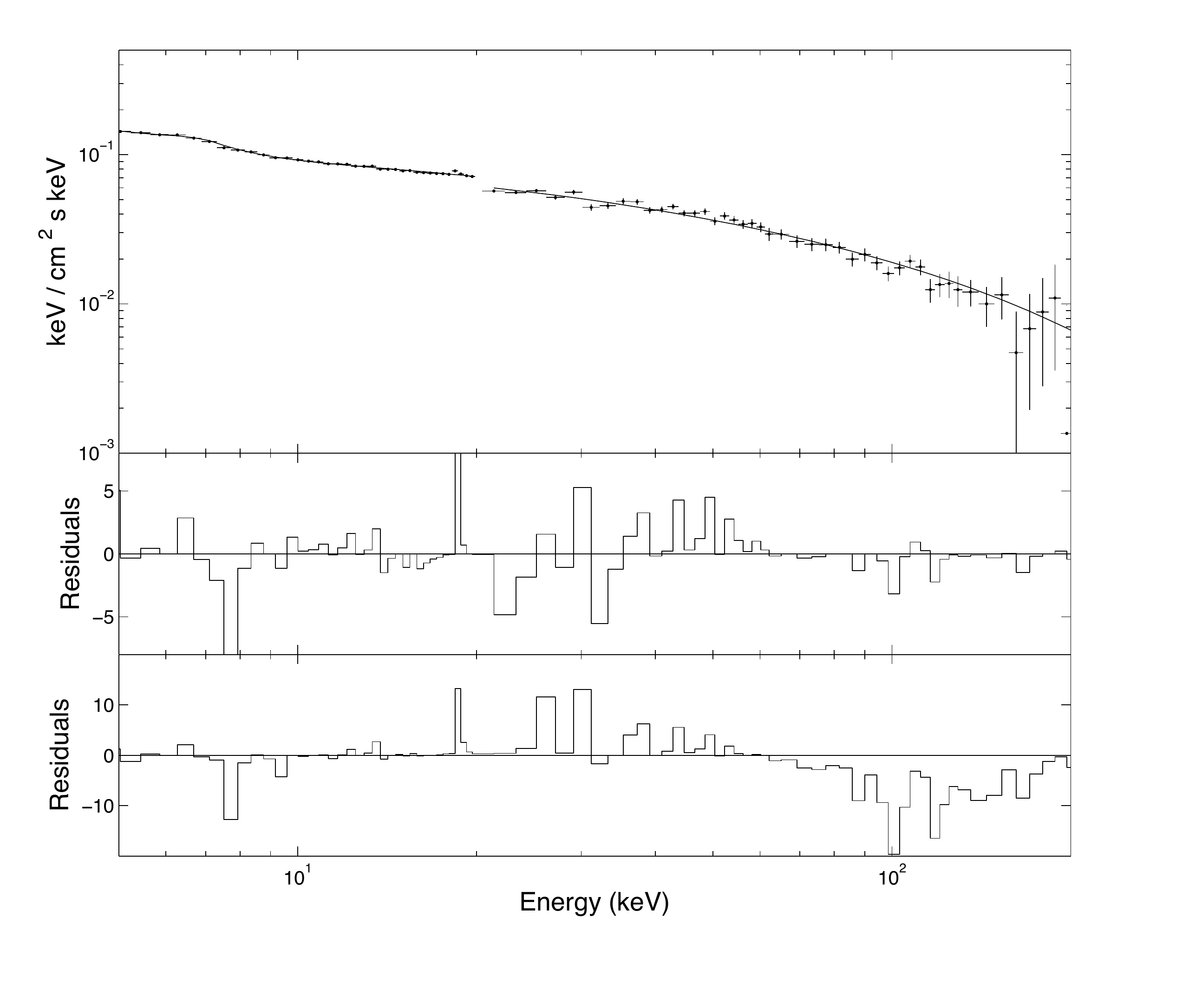}
\includegraphics[height= 7.4 cm,width= 8.50cm]{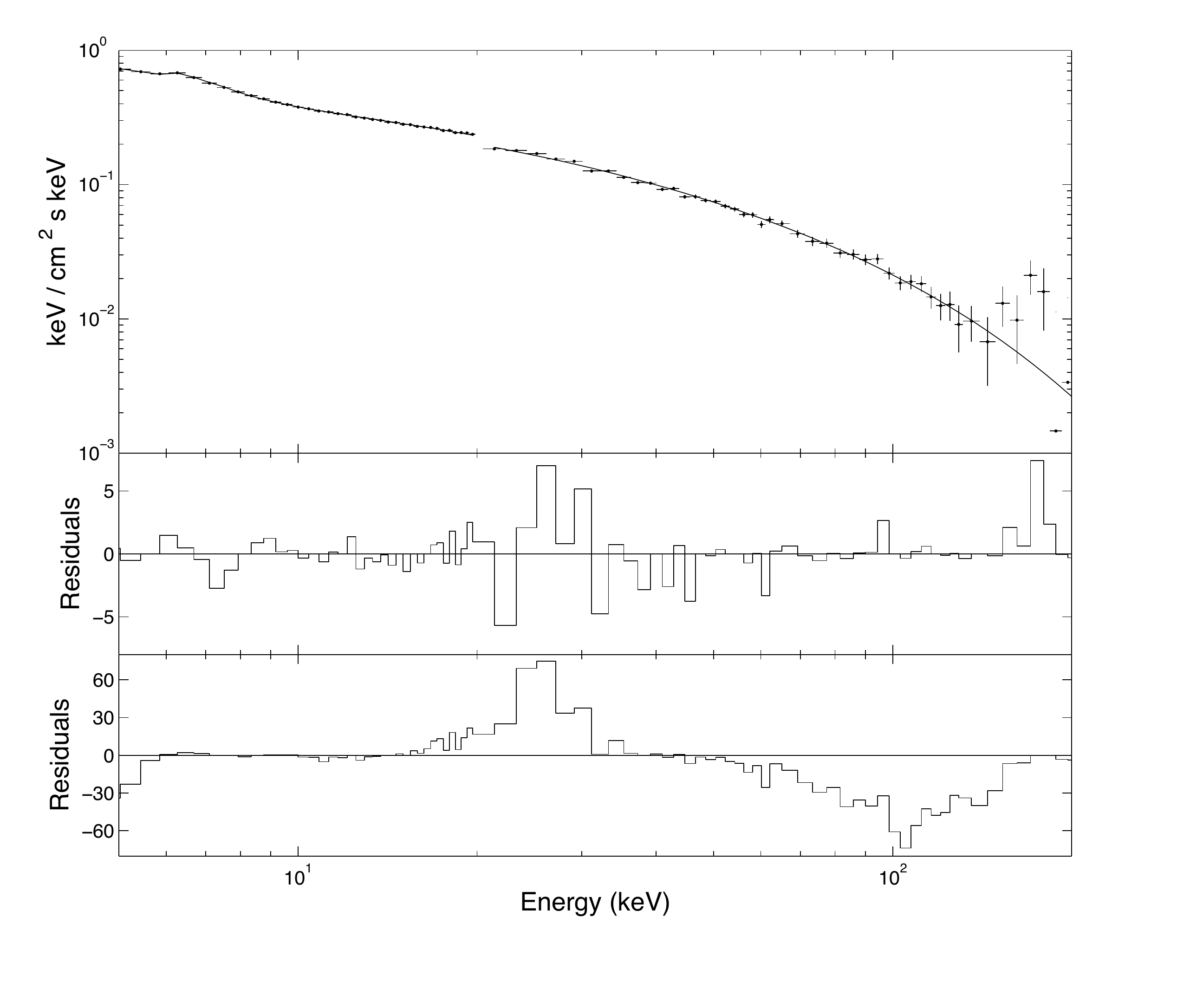}
\caption{Spectral fit results of the combined PCA and HEXTE spectra of GX 339-4 for two selected observations. For both spectra we used a model consisting of an interstellar absorption component, a gaussian line, a smeared edge in order to account for the reflection component, a multicolor disk-blackbody and a cutoff powerlaw. Top panel: observation \#1 (LHS, right vertical branch in the HID. \ref{fig:HID}). This observation shows the highest value of the cutoff we can consider reliable (small errors). The high values we found during the HSS presents too large error bands to be considered reliable.  Bottom panel: observation \#19 (LHS, upper part of the right vertical branch in the HID) shows one of the lowest values of the high-energy exponential cutoff we observed in our data.  The top panels in both the figures show a spectrum fitted with a cut-off powerlaw model (see text), the middle panels show residuals from a fit with a simple cutoff-powerlaw model, the bottom panels show the residuals for a simple powerlaw model. }\label{fig:spettri}
\end{figure}

In order to fit spectra we started trying a model of only one component, either cutoff power law or disk, but it could not fit all the spectra. A combination of the two was successful, with the exception of a few observations where a single component was sufficient. 
A simple model consisting of a multi-color disc-blackbody ({\tt diskbb})
and a cut-off power law ({\tt cutoffpl}) was used to fit spectra. A
hydrogen column density measured with instruments having a
low-energy coverage, e.g. Chandra, was taken into account by
adding a {\tt wabs} component into XSPEC, with N$_{\rm H}$ frozen to $5 \times 10^{21}
{\rm cm}^{-2}$ (M\' endez $\&$ van der Klis 1997; Kong et al. 2000). An
iron emission line with centroid fixed at 6.4 keV was further needed in
order to obtain acceptable fits. The line never becomes wider than 0.8 keV\footnote{Miller et al. (2004) analysed spectra of GX 339-4 obtained though simultaneous XMM-Newton/EPIC-pn and Rossi XTE observations during a brigh phase of the 2002/2003 outburst. They revealed an extremely skewed, relativistic Fe K$\alpha$ emission line in the spectra with strong red wings and intrinsically broad due to the Doppler shift  near the innermost stable orbit.}. To account for cross-calibration problems,
a variable multiplicative constant for the HEXTE spectra (as
compared to the PCA) was added to the fits. In order to account
for a reflection component, we introduced a smeared edge with
energy between 7.1 and 9.3 keV. This component is alway $\sim$ 10 keV wide and does not vary during the source evolution, so we can assert that the smeared edge component does not affect the properties of the cutoff. For the first part of the outburst (from $\#1$ to $\#23$), a disk component was not needed in order to obtain good spectral fits and was therefore removed from the fit. The average reduced $\chi^2$ was 1.17 for 92 degrees of freedom. In Fig. \ref{fig:spettri} spectral fits from selected observations are shown. The upper spectrum is from observation \#1 (belonging to the LHS, in the right vertical branch of the HID in Fig. \ref{fig:HID}).  The high-energy cutoff for this spectrum is the highest seen that can be considered reliable. Several HSS observations present very high values for the high-energy cutoff, but with very large uncertainties. The lower spectrum is from observation \#19 (belonging to the LHS, in the  right vertical branch very high part in the HID). This spectrum shows one of the lowest  cutoff energies we observed in our data. This observation was taken just before the LHS-to-HIMS transition.
For the second group of observations  (from $\#24$ to \#32, from \#36 to \#39 and from \#43 to \#50), a disk black body was necessary, yielding an average reduced $\chi ^2$ of
0.90 for 90 degrees of freedom. 
Finally, the softest observations (from $\#33$ to $\#35$, from $\#40$ to $\#42$ and from \#50 to \#83) 
did not require a high-energy cutoff. The average reduced $\chi^2$ was 0.92 for 91 degrees of freedom.

\begin{figure}
   \centering
   \includegraphics[height= 11.27cm,width= 8.5cm]{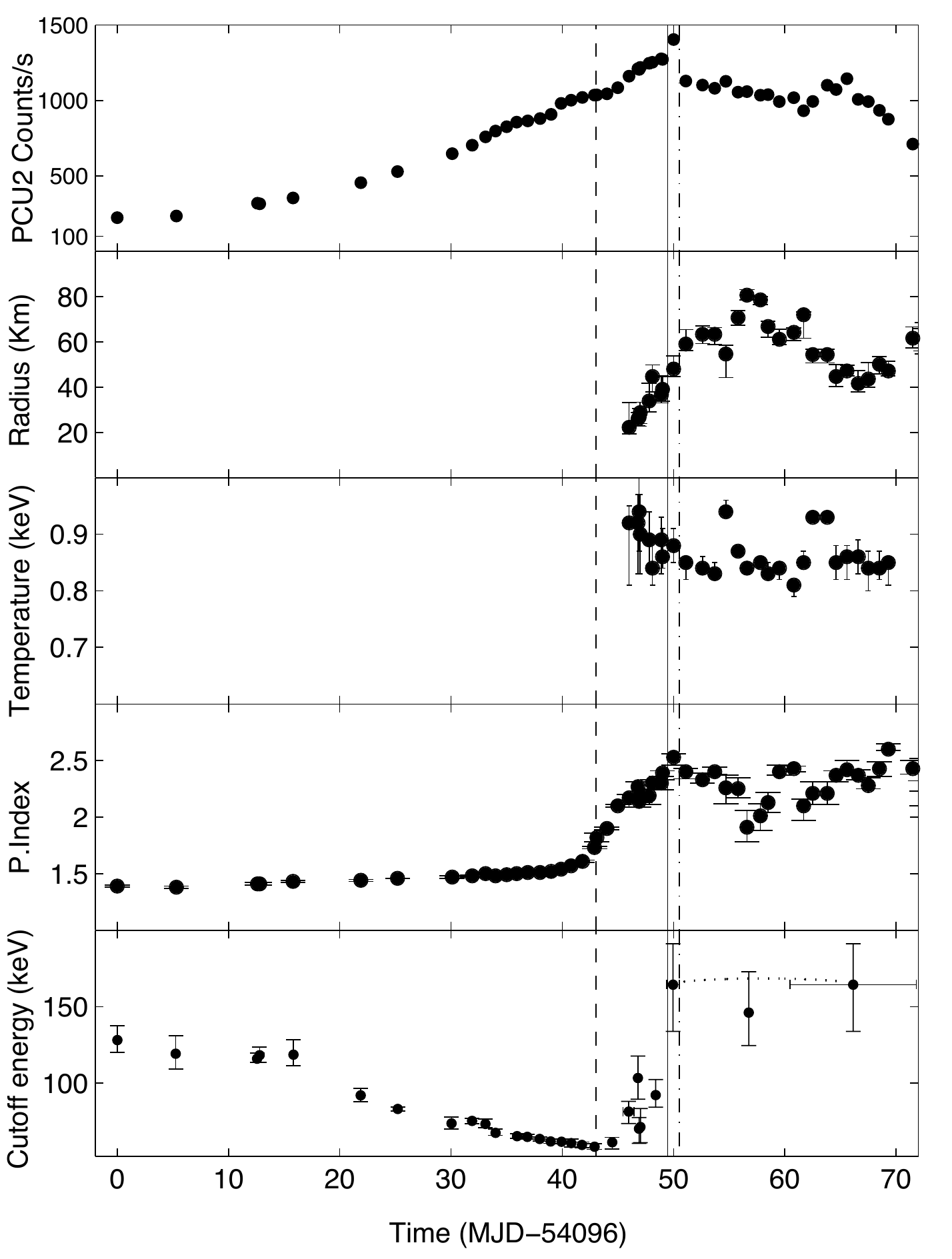}
   \caption{Light curve and evolution of the main spectral parameters of the source. For all the panels shown we plotted the spectral parameters for a selection of 51 observation (from \#1 to \#51) covering the entire hard-to-soft transition of the source, coming from Tab. \ref{tab:spettrali}. From the top to bottom: Light curve, inner radius in km (assuming a distance of 8 kpc and inclination of 60 degrees), disc temperature in keV at the inner radius, photon index, cutoff energy in keV. The vertical lines mark the transitions and follow the same convention as in Fig. \ref{fig:licu}. Points with horizontal error band correspond spectra obtained averaging observations with similar hardness. The error bar represents the time interval corresponding to the accumulation. In the bottom panel we used the values coming from Table \ref{tab:parametri_sommati}.  
   }
   \label{fig:parametri}
\end{figure}

\begin{figure}
   \centering
   \includegraphics[height= 5.56 cm,width= 9.3cm]{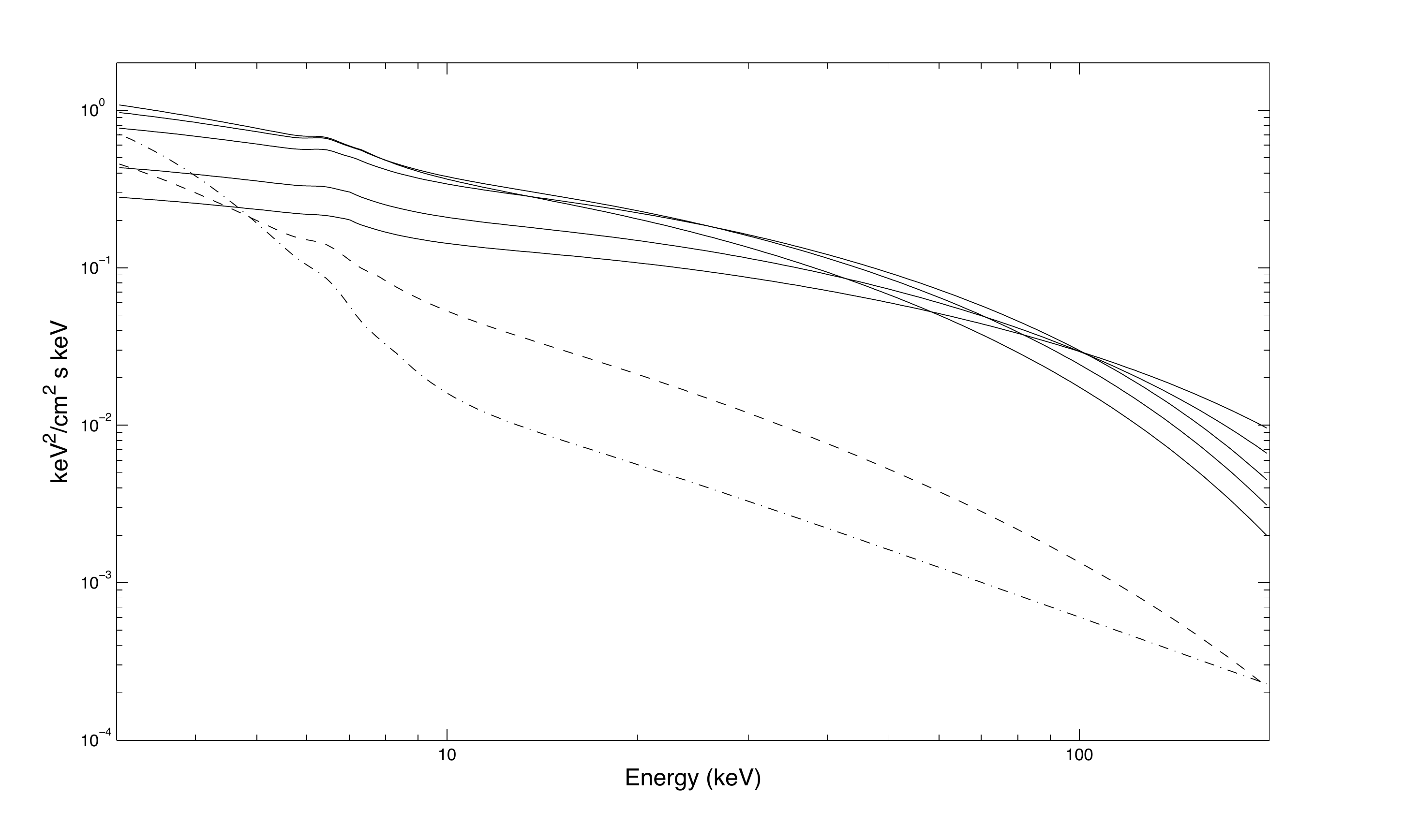}
   \caption{Different models fitting the Spectra in the LHS, HIMS and HSS. The solid lines correspond to LHS observations (\#5, \#7, \#15, \#19, \#20), the dashed line corresponds to a HIMS observation (\#25) and  the dotted dashed line corresponds to an HSS observation (\#33). The bottom two curves are scaled by a factor of 20 for clarity.}\label{fig:modello}
\end{figure}

The best fit parameters are listed in Tab. \ref{tab:spettrali}, while in Fig. \ref{fig:parametri} one can see the evolution of the main spectral components. The data shown in all the panels in Fig. \ref{fig:parametri} come from a selection of 51 observations (from \#1 to \#51) covering the entire hard-to-soft transition of the source. 
Fig. \ref{fig:modello} shows the evolution of the best fit model.

In order to accumulate spectra with better statistics for soft
observations where a high-energy cutoff is not measured, we summed spectra corresponding
to similar hardness. The resulting fit parameters are shown in Tab. \ref{tab:parametri_sommati}: a high-energy cutoff is detected, although with large uncertainties. 
In Fig. \ref{fig:parametri}, where we also mark the transitions established through timing analysis, we can follow the evolution of spectral prameters. In the bottom panel of Fig. \ref{fig:parametri} we used the parameters coming from Tab. \ref{tab:parametri_sommati}, while for the other panels of Fig. \ref{fig:parametri} we used parameters coming from Tab. \ref{tab:spettrali}. 
From Fig. \ref{fig:parametri} we can see that:

\begin{itemize}
\item The photon index undergoes a slight increase during the LHS from $\Gamma \sim 1.4$ to $\sim 1.6$. During the HIMS it rises from 1.6 to 2.4 in a few days. After the HIMS to SIMS transition the photon index is consistent with being constant. 
\item The high-energy cut off is clearly present in the LHS, during which it changes almost monotonically from $\sim$120 keV to $\sim$60 keV. After the transition to the HIMS  the high-energy cut off increases considerably, from 60 keV to 100 in a few days. During the SIMS and the HSS, the high-energy cut off appears to be very high ($\sim$160 keV) and constant, while after the final transition to the HSS (taking place between observations \#50 and \#51) it disappears. Since in this phase of the outburst the hard tail of the spectrum is very weak and a small fluctuation in the flux can modify  the spectrum, during the SIMS and the HSS the high-energy cutoff appears with large uncertainties. Therefore, we cannot assert the presence or the absence of a high-energy cutoff.  The fact that this happens near the HIMS-SIMS and SIMS-HSS transitions is particularly relevant in the context of  jets models, because the high-energy cutoff changes take place very close in time to the moment in which the jet emission is supposed to happen (Fender et al 2009).

In the first part of the outburst (corresponding roughly to the LHS), there is no measurable thermal disk component. When the disk appears, after the LHS-HIMS transition, we observe an
increase in the disk radius through the HIMS-SIMS-HSS transitions, followed by a decrease. At the same time, the disk temperature decreases steadily. Given the simplified form of our model, the absolute measurements of the inner radius are not robust and therefore not reliable. While the disk component was needed it could not be constrained very well and that we cannot draw any firm conclusions about its behavior from our fits. In addition, the poow low-energy sensitivity of RXTE usually makes very difficult the measurement of the spectral parameters related to the low-energy components.
\end{itemize}


\section{Discussion}\label{sec:discussion}


From the results persented in the previous section, we conclude that we observed three different
 transitions: from LHS to HIMS, from HIMS to SIMS and from SIMS to HSS. 

\begin{itemize}
\item The LHS-HIMS transition took place between Obs \#19 and Obs. \#20. It is identified through the appearance of a stronger type-C QPO in the power density
spectra, by a change in the parameters of the hard spectral component, and by a large change in the hardness (see Fig. \ref{fig:HID}).  The power-law
index increases faster with time across this transition, while the high-energy cut off stops decreasing. 

\item The first HIMS-SIMS transition took place between Obs. \#31 and Obs. \#32 and, by definition, was marked by the disappearance of the type-C QPO typical of the LHS and HIMS and the onset of a
type-B QPO in Obs. \#32. This observation is one of the four observations in the SIMS, all of them showing a type-A or type-B QPO. At the same time the high-energy cutoff shows a large change, jumping from $\sim$150 keV to higher values.

\item  The first SIMS-HSS transition takes place between Obs. \#32 and Obs. \#33, identified through the low value of integrated fractional rms and the absence of type-A/B/C QPOs. 

\item Besides these three main transition we observed transitions involving the intermediates states. 

To summarize, we observed:
\begin{itemize}
\item a transition from LHS to HIMS
\item a main transition and two secondary transitions from HIMS to SIMS
\item a main and a secondary transition from SIMS to HSS (this last transition is reported in Del Santo et al. 2008)
\end{itemize}
\end{itemize}

In Fig. \ref{fig:licu}, \ref{fig:HID}, \ref{fig:parametri}, we marked the LHS-HIMS transitions and only the first transitions from HIMS to SIMS and from SIMS to HSS. We refer to these first transition as the main transitions. 
Due to the short time scale of the transitions to and from the SIMS, we cannot be sure that we observed all transitions underwent by the source. All we can say is that the line in the HID corresponding to the HIMS-SIMS transition was crossed at least five times in total. The location of the three main transitions in the HID (see Fig. \ref{fig:HID}) was consistent with those of the previous outbursts. 


The observed behaviour of the photon index is also similar to that of the previous outbursts (Belloni et al. 2005; Del Santo et al. 2008), while the high-energy cut off behavior shows different properties. In 2004, the major HIMS to SIMS transition on the primary horizontal branch was simultaneously observed with INTEGRAL and RXTE (Belloni et al. 2006): after the transition, these authors report the lack of the high energy cut-off in the SIMS (present at $\sim$70 keV in the HIMS). Del Santo et al. (2008) confirm the latter result (disappearance of the cut-off in the SIMS) for the 2004 outburst by using simultaneous IBIS, SPI and JEM-X data collected during the same transition. However, they found a higher value of the cut-off in that same HIMS ($115_{-23}^{+27}$ keV) because of new INTAGRAL calibrations. 
Del Santo et al. (2009), using RXTE/PCA, RXTE/HEXTE and INTEGRAL/IBIS/ISGRI data performed a broad band spectral analysis covering the energy range from 3 keV (PCA) to 200 keV (IBIS/ISGRI). They observed a secondary HIMS/SIMS transition in the 2006/2007 outburst of GX 339-4 and found a different  behaviour in relation to what we observed: during the transition HIMS/SIMS the high-energy cut-off has moved to a lower energy, while we observed an opposite trend. Moreover, this variation is observed to take place {\it before} the HIMS/SIMS transition as deduced from the timing properties. Thanks to our RXTE data we clearly observed a high-energy cut off increase during the HIMS to SIMS transition. 

Miyakawa et al. (2008) studied a large sample of RXTE data from GX 339-4 during the hard state through different outbursts (not including the one presented here) using the same model we adopted (powerlaw with high energy exponential cutoff, smeared edge, Fe-line component). They found that a cutoff energy is present in all their hard-state observations. They could not make a real comparisons with previous results (Zdziarski et al. 1998), because those authors used different models, but  they clearly found a variable energy cut-off in GX 339-4, with values between 40 and 200 keV or more, similar to our results. The power law photon index showed an anticorrelation with the source luminosity. The cut off energy was also anticorrelated with luminosity above $10^{37}$ erg s$^{-1}$, while it was constant around $\sim$200 keV below that value.
Our LHS data span luminosities from $10^{37}$ erg s$^{-1}$  to $10^{38}$ erg s$^{-1}$. We observe the same anticorrelation cut off-luminosity, which continues in the HIMS up to $2.6 \times 10^{38}$erg   s$^{-1}$. 

They also calculated the absorption-corrected X-ray luminosities in the 2-200 keV range and assumed the distance of GX 339-4 to be 8 kpc (Zdziarski et al. 2004).  They obtained luminosities ranging from $1.0 \times 10^{37}$ erg s$^{-1}$ to $2.1 \times 10^{38}$ erg s$^{-1}$ and they found a clear anti-correlation between luminosity and the cutoff energy for luminosities  $> 7 \times 10^{37} $ erg s$^{-1}$. On the other hand they observed that the value of the high-energy cutoff seemed to be roughly constant at 200 keV when the luminosity was $< 7 \times 10^{37} erg s^{-1} $. 
We calculated the absorption-corrected luminosity of the source  following the same criteria Miyakawa et al. (2008) used and we found a similar range of luminosity (from $1.01 \times 10^{38}$ erg s$^{-1}$  to $1.14 \times 10^{37}$ erg s${-1}$). We found the same anticorrelation during the LHS and during the HIMS, while the luminosity ranged from $1.01 \times 10^{38}$ erg s$^{-1}$ and $2.57 \times 10^{38}$ erg s$^{-1}$. During the HSS the luminosity decreases from $7.25 \times 10^{37}$ erg s$^{-1}$ to $1.14 \times 10^{37}$ erg s$^{-1}$. As we pointed out before, during the SIMS and the HSS after the final transition from SIMS to HSS, we cannot exclude the presence of a cutoff, that either remains constant in energy or disappears.  This behavior is consistent with a constant cutoff around 200 keV with a luminosity $< 7 \times 10^{37}$erg s$^{-1} $.



\begin{figure}
\begin{center}
\includegraphics[height= 6.52 cm,width= 9.3 cm]{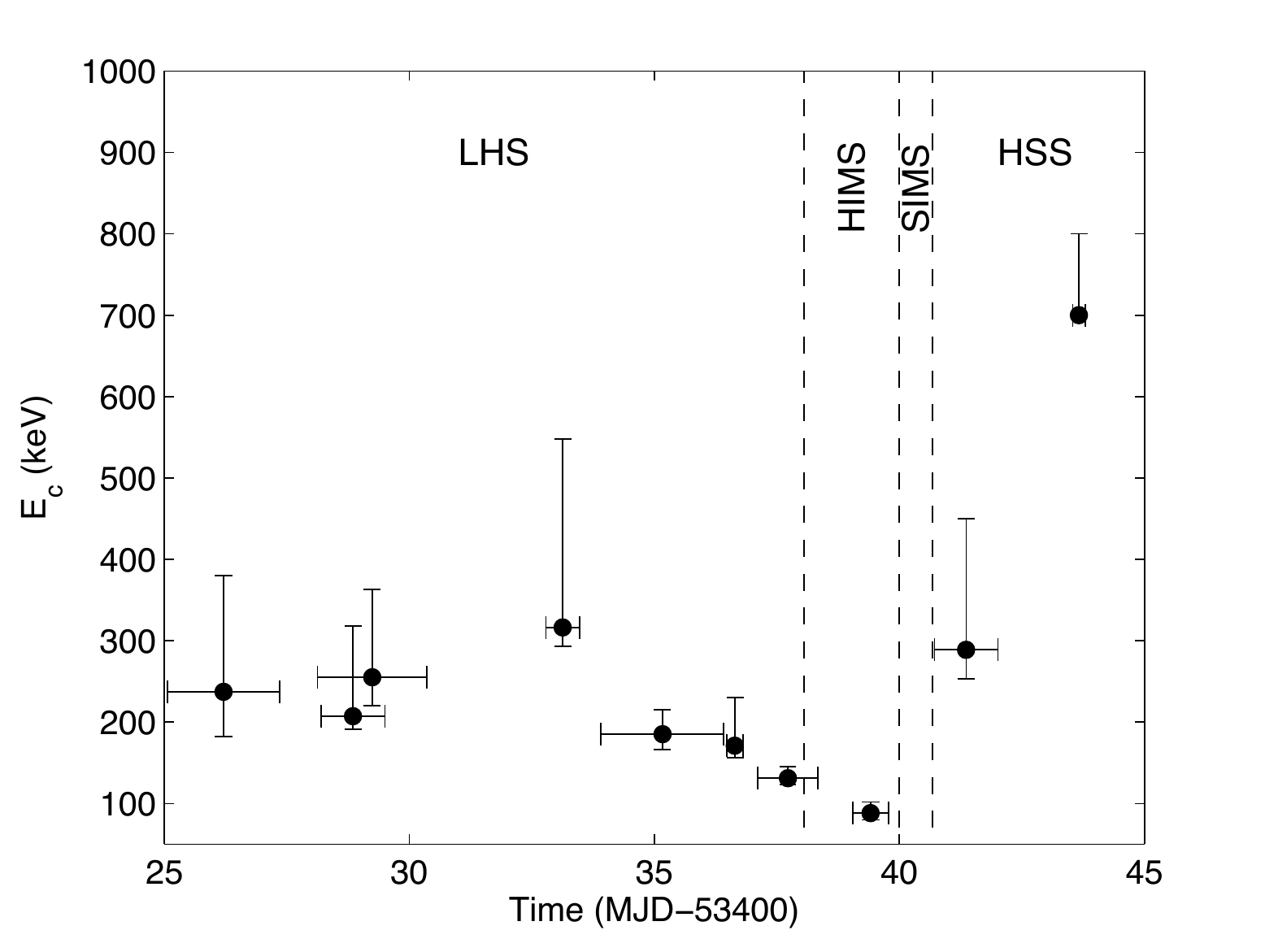}
\caption{Time evolution of the high-energy cutoff for the 2005 outburst of GRO J1655-40 as measured by INTEGRAL (from the data of Joinet, Kalemci, \& Senziani 2008), with RXTE-determined states. }\label{fig:cutoff_1655}
\end{center}
\end{figure}

The observed behaviour of the high energy cutoff of GX 339-4 is also similar to that observed with RXTE-INTEGRAL-Swift during the 2005 outburst of GRO J1655-40 (Joinet, Kalemci \& Senziani 2008), in contrast to the results of Caballero Garc\'ia et al. (2006) who do not find evidence of a cut off in the INTEGRAL spectra of the same source in the LHS. From their table, it is possible to reconstruct the time evolution of the high-energy cutoff, shown in Fig. \ref{fig:cutoff_1655}. 
They used various models available in the standard XSPEC 11.3.1 fitting package.  For all models the iron emission line, a multicolor disk blackbody and an interstellar absorption component were present. They first  fitted the data with a reflection model (PEXRAV in XSPEC) consisting of a power-law with  a high-energy cutoff and reflection from neutral medium. Since the hard powerlaw plus cutoff model in the LHS is usually interpreted as thermal Comptonization in a hot optically thin plasma, they also used the COMPTT model (Titarchuk 1994) in order to describe the high-energy spectrum. They observed that, with both models, the cutoff decreases through the LHS and the HIMS, starting from $\sim 200$ keV and reaching $\sim 60$ keV, to increase again in SIMS and HSS. The coverage of this outburst is good but unfortunately the HIMS was particularly short. 

The high-energy cutoff appears to be changing much more rapidly than other spectral parameters and possibly as fast as the timing properties. From Fig. \ref{fig:HID} and Fig. \ref{fig:parametri} (first panel from the bottom) we can see that transition from LHS to HIMS is evident both in the HID and in the cutoff evolution, while we cannot say the same thing for the HIMS-to-SIMS transition. The cutoff energy shows  a big variation crossing the HIMS-to-SIMS line, jumping from $\sim$100 keV to $\sim$160 keV, and at the same time the PDSs change very quickly. 
It is known that the ejection of transient relativistic jets typical of most black-hole binaries (see Fender, Belloni, Gallo 2004) takes place on very short time scales. The variations of the high-energy cutoff takes place on comparable timescales. Recently a comparative study of different systems has shown (crossing of the "jet line") that the jet ejection and HIMS/SIMS state transitions are not exactly simultaneous (Fender, Homan, Belloni 2009). Our data make possible to assert that the variation of the high-energy cutoff takes place just in correspondence with the main HIMS/SIMS transition, but due to the lack of radio observations we cannot exclude that the jet line coincides to the transition. In other words, it is possible that changes in the high-energy part of the spectrum and the crossing of the jet line are always simultaneous.     
Clearly the idea of a ``canonical" 100 keV cutoff (see  Zdziarski at al. 1996) in the LHS of black-holes binaries is too simplified. Large variations are seen in at least two sources across the LHS. These are transient systems, but also for persistent sources such as Cyg X-1 this paradigm needs to be revised (see Wilms et al. 2006).

The transition can also be seen in the photon index evolution (see Fig. \ref{fig:parametri}, fourth panel from the top). Even though the evolution of the photon index across the transitions seem to be continuous we clearly see a change in its trend. 
Summarizing, in coincidence to the HIMS/SIMS transition, defined through timing properties, we see: a sudden increase n the high-energy cutoff and a reverse in the trend of the powerlaw slope. We cannot exclude that these spectral changes are directly associated to the ejection of transient relativistic jets. 

Qualitatively, we can understand the reason for the softening in the LHS. Independent of whether the inner disk of the accretion disk moves inward or not, as the source becomes brighter more soft photons will be emitted by the disk. The photon input to the Comptonizing medium will therefore increase. This will steepen the power-law part of the Comptonization spectrum and will cool the population of electrons (see Sunyaev \& Titarchuck, 1980).
On the contrary, during the HIMS, when the softening is much more marked and the thermal disk starts to dominate, the increase in cutoff energy and hence in temperature of the Comptonizing cloud can not be explained within this framework.

The softening can therefore be understood in terms of thermal comptonization, but not the subsequent evolution. The idea that two varying powerlaw components, one associated with a high-energy cutoff (thermal) and one without it (non-thermal), cannot explain the high-energy cutoff evolution after the LHS to HIMS transition. The behaviour seems to indicate that it is only one component that evolves. Therefore,  a simple disappearing of a thermal component and its replacement with a non-thermal one is not a favored scenario. 
However,  Del Santo et al. (2008) show that the HIMS spectra from INTEGRAL observations indicate the presence of an additional component to the thermal compton one, evidence of the presence of a non-thermal tail in the distribution of the electrons. This tail could become dominant when approaching  the soft state, mimicking an increase in cutoff energy.
We tried to simulate a spectrum from a model consisting of a cutoff-powerlaw (with the parameters found at the LHS/HIMS transition) plus a simple powerlaw  (with the parameters corresponding to those of the SIMS and decreasing fluxes). We then fitted the simulated spectra with a cutoff powerlaw: we found that the high-energy cutoff does not vary significantly in response to the powerlaw addition. Therefore we can conclude that the  non-thermal powerlaw component, if present, does not influence the high-energy cutoff evolution.

\section{Conclusions}

The results presented above constitute an important measurement of the changes of the broad-band X-ray spectrum of a BHT across the hard to soft state transition, which is necessary for the development and testing of theoretical models. We have presented RXTE observations of GX 339-4 which covered the first half of the outburst. We followed the source spectral evolution from the LHS through the HIMS and the SIMS  untill the HSS. Our detailed broadband spectral analysis showed that the hard spectral component steepens during the transition from the hard to the soft state and the high-energy cutoff varies non-monotonically and rapidly though the transitions. The high-energy cutoff decreased monotonically from 120 to 60 keV during the brightening of the hard state, but increased again to 100 keV during the softening in the hard intermediate state. In the short-lived soft intermediate state the cutoff energy was $\sim$ 130 keV, but was no longer detected in the soft state. The changes in the high-energy cutoff were interpreted as a conseguence of the transitions. The high-energy cutoff behavior is similar to what observed with RXTE-INTEGRAL-Swift during the 2005 outburst of GRO J1655-40. The transitions can also be seen in the photon index evolution (see Fig. \ref{fig:parametri}, fourth panel from the top). Even though the evolution of the photon index across the transitions seem to be continuous we clearly see a change in its trend. From our analysis it is clear that although the transition from the LHS to the HSS is a process that takes days to weeks (see e.g. Belloni et al. 2005), a sharp transition in the properties of fast time variability takes place on a much shorter time scales, similarly to what happens to the high-energy cutoff. From what we have shown  it is clear that the transition in the properties of fast time variability corresponds also to a change in the high-energy properties.


\section*{acknowledgements}

This work has been supported by the Italian Space Agency through grants I/008/07/0 and I/088/07/0. TB acknowledges support from the International Space Science Institute

\newpage

\begin{figure}
\begin{center}
\includegraphics[height= 6 cm,width= 8.5 cm]{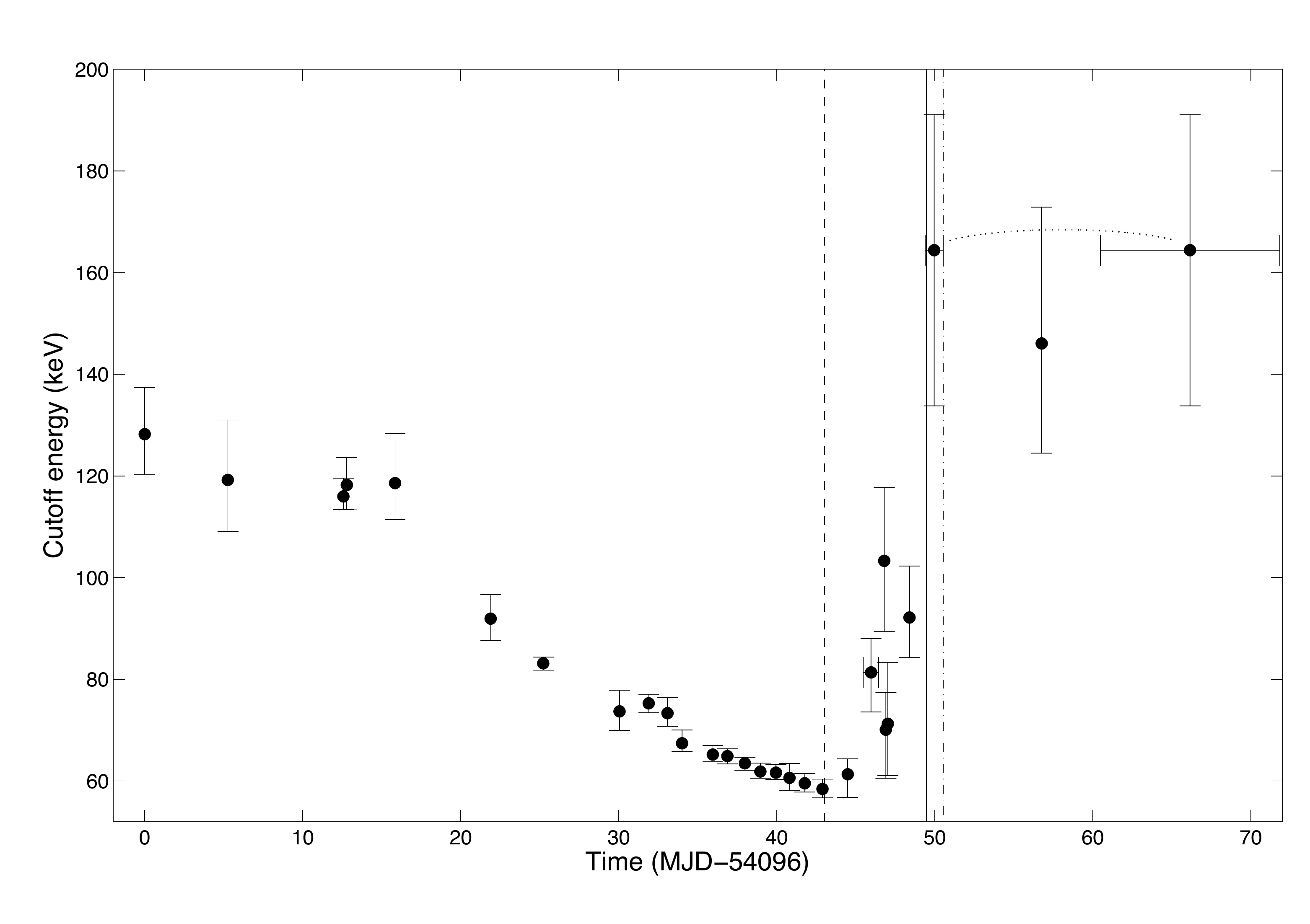}
\caption{Time evolution of the high-energy cutoff in GX 339-4 during its 2006/2007 outburst. Notice that after MJD 54165 (observation \#51) there is no measurable cutoff. The four vertical lines mark the LHS-HIMS (dashed), HIMS-SIMS(filled), SIMS-HSS (dot-dash) transitions. The dotted curve indicates that those two points belong to the same spectral fit. The points with horizontal error band corresponds to spectra summed across multiple observations. This figure is the same of the bottom panel of figure \ref{fig:parametri}.}\label{fig:cutoff}
\end{center}
\end{figure}
																																				
\begin{table*}																													
\renewcommand{\arraystretch}{1.3}																													
\begin{center}																													
\begin{tabular}{|c|c|c|c|c|c|c|c|c}																													
\hline																													
\#	&	Obs. ID	&	MJD	&		PCU2 Counts/s		&				hardness				&		rms				&	QPO	&	State (B)	&	State (M/R)	\\
\hline																													
\hline

1	&	92052-07-04-00	&	54096.00	&	$	223.50	\pm	0.38	$	&	$	0.885	\pm	0.004	$	&	$	47.93	\pm	2.81	$	&		&	LHS	&	INT	\\
2	&	92052-07-05-00	&	54101.26	&	$	234.30	\pm	0.50	$	&	$	0.882	\pm	0.005	$	&	$	45.09	\pm	7.11	$	&		&	LHS	&	INT	\\
3	&	92428-01-01-00	&	54108.57	&	$	318.70	\pm	0.43	$	&	$	0.873	\pm	0.003	$	&	$	45.66	\pm	2.76	$	&		&	LHS	&	H	\\
4	&	92428-01-01-01	&	54108.79	&	$	314.80	\pm	0.46	$	&	$	0.871	\pm	0.003	$	&	$	48.96	\pm	3.76	$	&		&	LHS	&	H	\\
5	&	92052-07-06-00	&	54111.84	&	$	354.10	\pm	0.63	$	&	$	0.868	\pm	0.004	$	&	$	43.68	\pm	2.85	$	&		&	LHS	&	H	\\
6	&	92052-07-06-01	&	54117.89	&	$	455.60	\pm	0.75	$	&	$	0.851	\pm	0.003	$	&	$	39.48	\pm	3.20	$	&		&	LHS	&	H	\\
7	&	92428-01-02-00	&	54121.21	&	$	527.70	\pm	0.70	$	&	$	0.847	\pm	0.003	$	&	$	38.82	\pm	1.08	$	&		&	LHS	&	H	\\
8	&	92428-01-03-00	&	54126.05	&	$	647.40	\pm	1.15	$	&	$	0.833	\pm	0.004	$	&	$	36.43	\pm	2.69	$	&		&	LHS	&	H	\\
9	&	92035-01-01-01	&	54127.89	&	$	703.70	\pm	0.98	$	&	$	0.837	\pm	0.003	$	&	$	36.65	\pm	2.49	$	&	C	&	LHS	&	H	\\
10	&	92035-01-01-02	&	54129.08	&	$	758.40	\pm	1.20	$	&	$	0.831	\pm	0.003	$	&	$	36.41	\pm	3.51	$	&	C	&	LHS	&	H	\\
11	&	 92035-01-01-02 	&	54129.08	&	$	797.40	\pm	1.07	$	&	$	0.832	\pm	0.003	$	&	$	35.87	\pm	1.09	$	&	C	&	LHS	&	H	\\
12	&	92035-01-01-04	&	54131.03	&	$	826.50	\pm	1.12	$	&	$	0.830	\pm	0.003	$	&	$	35.08	\pm	1.06	$	&	C	&	LHS	&	H	\\
13	&	92035-01-02-00	&	54131.95	&	$	856.20	\pm	1.15	$	&	$	0.826	\pm	0.003	$	&	$	33.33	\pm	1.81	$	&	C	&	LHS	&	H	\\
14	&	92035-01-02-01	&	54132.87	&	$	863.80	\pm	1.18	$	&	$	0.823	\pm	0.003	$	&	$	32.37	\pm	1.94	$	&	C	&	LHS	&	H	\\
15	&	92035-01-02-02	&	54133.98	&	$	879.50	\pm	1.18	$	&	$	0.823	\pm	0.003	$	&	$	32.14	\pm	0.94	$	&	C	&	LHS	&	H	\\
16	&	92035-01-02-03	&	54134.96	&	$	908.40	\pm	1.22	$	&	$	0.818	\pm	0.003	$	&	$	31.88	\pm	0.91	$	&	C	&	LHS	&	H	\\
17	&	92035-01-02-04	&	54135.94	&	$	981.00	\pm	1.32	$	&	$	0.810	\pm	0.003	$	&	$	30.93	\pm	1.66	$	&	C	&	LHS	&	H	\\
18	&	92035-01-02-08	&	54136.80	&	$	1001.00	\pm	1.63	$	&	$	0.801	\pm	0.003	$	&	$	28.62	\pm	2.81	$	&	C	&	LHS	&	H	\\
19	&	92035-01-02-07	&	54137.77	&	$	1020.00	\pm	1.42	$	&	$	0.782	\pm	0.003	$	&	$	26.20	\pm	1.54	$	&	C	&	LHS	&	H	\\
\hline																													
20	&	92035-01-02-06	&	54138.89	&	$	1035.00	\pm	1.44	$	&	$	0.732	\pm	0.002	$	&	$	22.72	\pm	1.18	$	&	C	&	HIMS	&	H	\\
21	&	92035-01-03-00	&	54139.15	&	$	1037.00	\pm	1.45	$	&	$	0.714	\pm	0.002	$	&	$	23.10	\pm	0.46	$	&	C	&	HIMS	&	H	\\
22	&	92035-01-03-01	&	54140.00	&	$	1045.00	\pm	1.50	$	&	$	0.658	\pm	0.002	$	&	$	21.02	\pm	0.60	$	&	C	&	HIMS	&	H	\\
23	&	92035-01-03-02	&	54140.98	&	$	1084.00	\pm	1.62	$	&	$	0.574	\pm	0.002	$	&	$	20.21	\pm	1.04	$	&	C	&	HIMS	&	H	\\
24	&	92035-01-03-03	&	54141.96	&	$	1161.00	\pm	1.84	$	&	$	0.480	\pm	0.002	$	&	$	19.09	\pm	0.53	$	&	C	&	HIMS	&	INT	\\
25	&	92428-01-04-00	&	54142.80	&	$	1210.00	\pm	1.99	$	&	$	0.426	\pm	0.001	$	&	$	18.70	\pm	0.53	$	&	C	&	HIMS	&	INT	\\
26	&	92428-01-04-01	&	54142.90	&	$	1206.00	\pm	2.06	$	&	$	0.434	\pm	0.002	$	&	$	18.05	\pm	1.24	$	&	C	&	HIMS	&	INT	\\
27	&	92428-01-04-02	&	54143.03	&	$	1218.00	\pm	2.12	$	&	$	0.439	\pm	0.002	$	&	$	17.66	\pm	1.35	$	&	C	&	HIMS	&	INT	\\
28	&	92428-01-04-03	&	54143.82	&	$	1247.00	\pm	2.22	$	&	$	0.393	\pm	0.001	$	&	$	17.35	\pm	0.72	$	&	C	&	HIMS	&	INT	\\
29	&	92035-01-03-05	&	54144.06	&	$	1253.00	\pm	2.18	$	&	$	0.355	\pm	0.001	$	&	$	16.25	\pm	0.51	$	&	C	&	HIMS	&	INT	\\
30	&	92428-01-04-04	&	54144.91	&	$	1276.00	\pm	2.30	$	&	$	0.364	\pm	0.001	$	&	$	16.33	\pm	0.69	$	&	C	&	HIMS	&	INT	\\
31	&	92035-01-03-06	&	54144.97	&	$	1272.00	\pm	2.19	$	&	$	0.365	\pm	0.001	$	&	$	16.24	\pm	0.46	$	&	C	&	HIMS	&	SPL	\\
\hline																													
32	&	92035-01-04-00	&	54145.96	&	$	1404.00	\pm	2.62	$	&	$	0.276	\pm	0.001	$	&	$	11.22	\pm	0.59	$	&	B	&	SIMS	&	SPL	\\
\hline																													
33	&	92035-01-04-01	&	54147.09	&	$	1129.00	\pm	2.43	$	&	$	0.180	\pm	0.001	$	&	$	7.98	\pm	0.69	$	&		&	HSS	&	INT	\\
34	&	92035-01-04-02	&	54148.64	&	$	1102.00	\pm	2.35	$	&	$	0.164	\pm	0.001	$	&	$	6.64	\pm	0.60	$	&		&	HSS	&	TD	\\
35	&	92085-01-01-00	&	54149.69	&	$	1080.00	\pm	2.29	$	&	$	0.166	\pm	0.001	$	&	$	6.21	\pm	0.58	$	&		&	HSS	&	INT	\\
36	&	92085-01-01-04	&	54150.75	&	$	1126.00	\pm	2.36	$	&	$	0.190	\pm	0.001	$	&	$	7.36	\pm	0.70	$	&		&	HSS	&	INT	\\
37	&	92085-01-01-05	&	54151.78	&	$	1055.00	\pm	2.28	$	&	$	0.152	\pm	0.001	$	&	$	6.65	\pm	0.59	$	&		&	HSS	&	TD	\\
38	&	92085-01-01-06	&	54152.63	&	$	1059.00	\pm	2.35	$	&	$	0.136	\pm	0.001	$	&	$	4.87	\pm	0.69	$	&		&	HSS	&	TD	\\
39	&	92085-01-02-00	&	54153.81	&	$	1034.00	\pm	2.29	$	&	$	0.136	\pm	0.001	$	&	$	6.74	\pm	0.56	$	&		&	HSS	&	TD	\\
40	&	92085-01-02-01	&	54154.52	&	$	1038.00	\pm	2.28	$	&	$	0.140	\pm	0.001	$	&	$	5.75	\pm	0.60	$	&		&	HSS	&	TD	\\
41	&	92085-01-02-02	&	54155.49	&	$	991.90	\pm	2.16	$	&	$	0.149	\pm	0.001	$	&	$	7.17	\pm	0.67	$	&		&	HSS	&	TD	\\
42	&	92085-01-02-03	&	54156.75	&	$	1018.00	\pm	2.12	$	&	$	0.185	\pm	0.001	$	&	$	6.85	\pm	0.55	$	&		&	HSS	&	INT	\\
43	&	92085-01-02-04	&	54157.73	&	$	931.50	\pm	2.06	$	&	$	0.137	\pm	0.001	$	&	$	7.47	\pm	0.66	$	&		&	HSS	&	TD	\\
44	&	92085-01-02-05	&	54158.44	&	$	993.80	\pm	2.09	$	&	$	0.183	\pm	0.001	$	&	$	7.31	\pm	0.62	$	&		&	HSS	&	INT	\\
45	&	92085-01-02-06	&	54159.84	&	$	1101.00	\pm	2.20	$	&	$	0.229	\pm	0.001	$	&	$	5.27	\pm	0.68	$	&	A	&	SIMS	&	SPL	\\
46	&	92085-01-03-00	&	54160.61	&	$	1072.00	\pm	1.96	$	&	$	0.305	\pm	0.001	$	&	$	15.10	\pm	0.53	$	&	C	&	HIMS	&	SPL	\\
47	&	92085-01-03-01	&	54161.60	&	$	1144.00	\pm	2.20	$	&	$	0.271	\pm	0.001	$	&	$	11.74	\pm	0.84	$	&	B	&	SIMS	&	SPL	\\

\hline																													
\end{tabular}																													
\end{center}																													
\end{table*}																													
\begin{table*}																													
\renewcommand{\arraystretch}{1.3}																													
\begin{center}																													
\begin{tabular}{|c|c|c|c|c|c|c|c|c}																													
\hline																													
\#	&	Obs. ID	&	MJD	&		PCU2 Counts/s				&		hardness				&		rms				&	QPO	&	State (B)	&	State (M/R)	\\
\hline																													
\hline																													
																													
48	&	92085-01-03-02	&	54162.64	&	$	1008.00	\pm	1.85	$	&	$	0.298	\pm	0.001	$	&	$	14.35	\pm	0.53	$	&	C	&	HIMS	&	SPL	\\
49	&	92085-01-03-03	&	54163.49	&	$	992.60	\pm	1.84	$	&	$	0.305	\pm	0.001	$	&	$	14.22	\pm	0.53	$	&	C	&	HIMS	&	SPL	\\
50	&	92085-01-03-04	&	54164.47	&	$	935.20	\pm	1.88	$	&	$	0.221	\pm	0.001	$	&	$	6.33	\pm	0.63	$	&	A	&	SIMS	&	SPL	\\
51	&	92085-01-03-05	&	54165.26	&	$	876.40	\pm	1.87	$	&	$	0.206	\pm	0.001	$	&	$	6.21	\pm	0.72	$	&		&	HSS	&	INT	\\
52	&	92085-01-04-13	&	54167.48	&	$	710.60	\pm	1.76	$	&	$	0.162	\pm	0.001	$	&	$	7.49	\pm	1.58	$	&		&	HSS	&	INT	\\
53	&	92085-01-04-08	&	54168.20	&	$	690.70	\pm	1.72	$	&	$	0.159	\pm	0.001	$	&	$	6.85	\pm	1.74	$	&		&	HSS	&	INT	\\
54	&	92085-01-04-09	&	54168.26	&	$	672.60	\pm	1.67	$	&	$	0.144	\pm	0.001	$	&	$	6.36	\pm	1.69	$	&		&	HSS	&	TD	\\
55	&	92085-01-04-10	&	54168.33	&	$	664.00	\pm	1.76	$	&	$	0.141	\pm	0.001	$	&	$	7.01	\pm	1.71	$	&		&	HSS	&	TD	\\
56	&	92085-01-04-00	&	54169.17	&	$	630.20	\pm	1.50	$	&	$	0.121	\pm	0.001	$	&	$	5.06	\pm	2.61	$	&		&	HSS	&	TD	\\
57	&	92085-01-04-02	&	54170.16	&	$	588.50	\pm	1.47	$	&	$	0.119	\pm	0.001	$	&	$	5.77	\pm	1.40	$	&		&	HSS	&	TD	\\
58	&	92085-01-04-03	&	54170.22	&	$	596.80	\pm	1.48	$	&	$	0.125	\pm	0.001	$	&	$	5.95	\pm	1.46	$	&		&	HSS	&	TD	\\
59	&	92085-01-04-05	&	54171.21	&	$	562.20	\pm	1.44	$	&	$	0.120	\pm	0.001	$	&	$	6.13	\pm	1.76	$	&		&	HSS	&	TD	\\
60	&	92085-01-04-06	&	54171.27	&	$	563.50	\pm	1.40	$	&	$	0.124	\pm	0.001	$	&	$	7.70	\pm	1.52	$	&		&	HSS	&	TD	\\
61	&	92085-01-04-07	&	54171.34	&	$	562.10	\pm	1.44	$	&	$	0.122	\pm	0.001	$	&	$	6.40	\pm	1.30	$	&		&	HSS	&	TD	\\
62	&	92085-01-04-04	&	54172.28	&	$	538.00	\pm	1.53	$	&	$	0.120	\pm	0.001	$	&	$	7.92	\pm	2.19	$	&		&	HSS	&	TD	\\
63	&	92085-01-04-11	&	54173.93	&	$	542.70	\pm	1.39	$	&	$	0.165	\pm	0.001	$	&	$	13.88	\pm	2.21	$	&		&	HSS	&	INT	\\
64	&	92085-02-01-01	&	54173.95	&	$	546.40	\pm	1.32	$	&	$	0.166	\pm	0.001	$	&	$	10.42	\pm	1.87	$	&		&	HSS	&	INT	\\
65	&	92085-02-01-00	&	54175.02	&	$	512.60	\pm	1.14	$	&	$	0.154	\pm	0.001	$	&	$	6.96	\pm	0.91	$	&		&	HSS	&	INT	\\
66	&	92085-02-01-02	&	54175.97	&	$	536.90	\pm	1.17	$	&	$	0.184	\pm	0.001	$	&	$	7.39	\pm	1.54	$	&		&	HSS	&	INT	\\
67	&	92085-02-01-03	&	54177.12	&	$	555.50	\pm	1.17	$	&	$	0.201	\pm	0.001	$	&	$	5.75	\pm	0.84	$	&		&	HSS	&	INT	\\
68	&	92085-02-01-04	&	54177.98	&	$	571.90	\pm	1.20	$	&	$	0.213	\pm	0.001	$	&	$	3.63	\pm	1.36	$	&		&	HSS	&	INT	\\
69	&	92085-02-01-05	&	54178.76	&	$	520.50	\pm	1.16	$	&	$	0.193	\pm	0.001	$	&	$	5.50	\pm	1.35	$	&		&	HSS	&	INT	\\
70	&	92085-02-01-06	&	54180.14	&	$	498.80	\pm	1.12	$	&	$	0.198	\pm	0.001	$	&	$	7.16	\pm	1.24	$	&		&	HSS	&	INT	\\
71	&	92085-02-02-03	&	54181.05	&	$	427.50	\pm	1.00	$	&	$	0.169	\pm	0.001	$	&	$	7.53	\pm	1.49	$	&		&	HSS	&	INT	\\
72	&	92085-02-02-00	&	54182.09	&	$	508.80	\pm	1.06	$	&	$	0.220	\pm	0.001	$	&	$	5.56	\pm	1.09	$	&		&	HSS	&	INT	\\
73	&	92085-02-02-02	&	54184.05	&	$	428.70	\pm	0.93	$	&	$	0.197	\pm	0.001	$	&	$	7.22	\pm	1.49	$	&		&	HSS	&	INT	\\
74	&	92085-02-02-01	&	54186.08	&	$	338.20	\pm	0.77	$	&	$	0.170	\pm	0.001	$	&	$	9.04	\pm	1.25	$	&		&	HSS	&	INT	\\
75	&	92085-02-03-00	&	54187.98	&	$	295.20	\pm	0.73	$	&	$	0.115	\pm	0.001	$	&	$	7.54	\pm	1.86	$	&		&	HSS	&	TD	\\
76	&	92085-02-03-01	&	54190.02	&	$	277.00	\pm	0.70	$	&	$	0.122	\pm	0.001	$	&	$	5.17	\pm	2.08	$	&		&	HSS	&	TD	\\
77	&	92085-02-03-02	&	54192.63	&	$	237.60	\pm	0.62	$	&	$	0.118	\pm	0.001	$	&	$	6.03	\pm	2.17	$	&		&	HSS	&	TD	\\
78	&	92085-02-04-00	&	54196.55	&	$	202.20	\pm	0.51	$	&	$	0.137	\pm	0.001	$	&	$	4.18	\pm	2.86	$	&		&	HSS	&	INT	\\
79	&	92085-02-04-02	&	54200.62	&	$	167.70	\pm	0.44	$	&	$	0.144	\pm	0.001	$	&	$	7.56	\pm	2.20	$	&		&	HSS	&	INT	\\
80	&	92085-02-05-00	&	54202.51	&	$	154.50	\pm	0.41	$	&	$	0.155	\pm	0.001	$	&	$	8.38	\pm	2.33	$	&		&	HSS	&	INT	\\
81	&	92085-02-05-01	&	54204.47	&	$	120.80	\pm	0.35	$	&	$	0.107	\pm	0.001	$	&	$	7.50	\pm	2.83	$	&		&	HSS	&	INT	\\
82	&	92085-02-05-02	&	54206.63	&	$	168.30	\pm	0.41	$	&	$	0.225	\pm	0.001	$	&	$	10.09	\pm	2.69	$	&		&	HSS	&	INT	\\
83	&	92085-02-05-03	&	54208.40	&	$	134.30	\pm	0.35	$	&	$	0.200	\pm	0.001	$	&	$	6.25	\pm	4.10	$	&		&	HSS	&	INT	\\
																													
\hline																													
\end{tabular}																													
\caption{The columns are: observation number in this work, RXTE observation ID, MJD, PCU2 count rate, hardness ratio, fractional rms (0.001 - 64 Hz), QPO type, and states according to Belloni (2009) and McClintock \& Remillard (2006).																													
}\label{tab:colori}																													
\end{center}																													
\end{table*}

\begin{table*}																																					
\renewcommand{\arraystretch}{1.3}																																					
\begin{center}																																					
\begin{tabular}{|c|c|c|c|c|}																																					
\hline																																					
$\#$	&	kT	(keV)								&		R	(Km)							&			$\Gamma$							&		E$_{c}$				\\
\hline																																					
\hline																																					

1	&		-								&		-								&	$	1.39	\pm	0.01					$	&	$	128.2	\pm	7.9	$	\\
2	&		-								&		-								&	$	1.38	\pm	0.01					$	&	$	119.2	\pm	10.1	$	\\
3	&		-								&		-								&	$	1.41	\pm	0.01					$	&	$	116.0	\pm	2.6	$	\\
4	&		-								&		-								&	$	1.41	\pm	0.01					$	&	$	118.2	\pm	4.9	$	\\
5	&		-								&		-								&	$	1.43	\pm	0.01					$	&	$	118.6	\pm	7.2	$	\\
6	&		-								&		-								&	$	1.44	\pm	0.01					$	&	$	91.9	\pm	4.3	$	\\
7	&		-								&		-								&	$	1.46	\pm	0.00					$	&	$	83.1	\pm	1.4	$	\\
8	&		-								&		-								&	$	1.47	\pm	0.01					$	&	$	73.7	\pm	3.7	$	\\
9	&		-								&		-								&	$	1.48	\pm	0.01					$	&	$	75.3	\pm	1.9	$	\\
10	&		-								&		-								&	$	1.50	\pm	0.01					$	&	$	73.3	\pm	2.6	$	\\
11	&		-								&										&	$	1.48	\pm	0.01					$	&	$	67.4	\pm	1.6	$	\\
12	&		-								&		-								&	$	1.49	\pm	0.01					$	&	$	65.8	\pm	1.3	$	\\
13	&		-								&		-								&	$	1.50	\pm	0.01					$	&	$	65.2	\pm	1.4	$	\\
14	&		-								&		-								&	$	1.51	\pm	0.01					$	&	$	64.9	\pm	1.6	$	\\
15	&		-								&		-								&	$	1.51	\pm	0.01					$	&	$	63.5	\pm	1.4	$	\\
16	&		-								&		-								&	$	1.52	\pm	0.01					$	&	$	61.8	\pm	1.3	$	\\
17	&		-								&		-								&	$	1.54	\pm	0.01					$	&	$	61.6	\pm	1.4	$	\\
18	&		-								&		-								&	$	1.57	\pm	0.01					$	&	$	60.6	\pm	2.6	$	\\
19	&		-								&		-								&	$	1.61	\pm	0.01					$	&	$	59.5	\pm	1.7	$	\\
\hline																																					
20	&		-								&		-								&	$	1.73	\pm	0.01					$	&	$	58.4	\pm	1.8	$	\\
21	&		-								&		-								&	$	1.82	\pm	0.04					$	&	$	79.6	\pm	15.3	$	\\
22	&		-								&		-								&	$	1.90	\pm	0.01					$	&	$	65.0	\pm	2.8	$	\\
23	&		-								&		-								&	$	2.10	\pm	0.01					$	&	$	80.4	\pm	4.9	$	\\
24	&	$	0.92	_{-	0.11	}	^{+	0.03	}	$	&	$	22.2	_{-	2.8	}	^{+	10.9	}	$	&	$	2.17	_{-	0.08	}	^{+	0.03	}	$	&	$	83.3	\pm	10.7	$	\\
25	&	$	0.92	_{-	0.09	}	^{+	0.05	}	$	&	$	26.0	_{-	3.0	}	^{+	2.7	}	$	&	$	2.27	_{-	0.06	}	^{+	0.04	}	$	&	$	104.8	\pm	22.9	$	\\
26	&	$	0.94	\pm	0.07					$	&	$	26.7	_{-	4.0	}	^{+	3.9	}	$	&	$	2.14	_{-	0.05	}	^{+	0.04	}	$	&	$	69.9	\pm	9.5	$	\\
27	&	$	0.90	\pm	0.07					$	&	$	28.7	_{-	4.8	}	^{+	4.6	}	$	&	$	2.18	_{-	0.05	}	^{+	0.02	}	$	&	$	71.8	\pm	11.0	$	\\
28	&	$	0.89	\pm	0.05					$	&	$	33.9	_{-	4.9	}	^{+	7.9	}	$	&	$	2.19	_{-	0.08	}	^{+	0.05	}	$	&	$	60.6	\pm	8.5	$	\\
29	&	$	0.84	_{-	0.03	}	^{+	0.05	}	$	&	$	44.6	_{-	6.8	}	^{+	5.2	}	$	&	$	2.30	_{-	0.05	}	^{+	0.03	}	$	&	$	95.3	\pm	16.0	$	\\
30	&	$	0.89	_{-	0.06	}	^{+	0.04	}	$	&	$	36.7	_{-	3.7	}	^{+	8.2	}	$	&	$	2.30	\pm	0.06					$	&	$	95.5	\pm	18.7	$	\\
31	&	$	0.86	_{-	0.02	}	^{+	0.05	}	$	&	$	39.1	_{-	5.4	}	^{+	5.6	}	$	&	$	2.39	_{-	0.06	}	^{+	0.02	}	$	&	$	156.6	\pm	42.0	$	\\
\hline																																					
32	&	$	0.88	\pm	0.03					$	&	$	48.0	_{-	3.4	}	^{+	5.8	}	$	&	$	2.53	_{-	0.07	}	^{+	0.03	}	$	&	$	254.5	\pm	111.9	$	\\
\hline																																					
33	&	$	0.85	_{-	0.03	}	^{+	0.01	}	$	&	$	59.0	_{-	2.8	}	^{+	6.4	}	$	&	$	2.40	_{-	0.06	}	^{+	0.03	}	$	&	$	-			$	\\
34	&	$	0.84	_{-	0.01	}	^{+	0.02	}	$	&	$	63.4	_{-	4.2	}	^{+	3.6	}	$	&	$	2.33	_{-	0.03	}	^{+	0.06	}	$	&	$	-			$	\\
35	&	$	0.83	_{-	0.01	}	^{+	0.02	}	$	&	$	63.4	_{-	4.7	}	^{+	2.9	}	$	&	$	2.40	_{-	0.02	}	^{+	0.04	}	$	&	$	-			$	\\
36	&	$	0.94	_{-	0.01	}	^{+	0.02	}	$	&	$	54.6	_{-	10.4	}	^{+	4.0	}	$	&	$	2.26	_{-	0.14	}	^{+	0.11	}	$	&	$	118.1	\pm	40.1	$	\\
37	&	$	0.87	\pm	0.01					$	&	$	70.7	_{-	3.4	}	^{+	3.2	}	$	&	$	2.25	_{-	0.08	}	^{+	0.10	}	$	&	$	163.4	\pm	70.8	$	\\
38	&	$	0.84	\pm	0.01					$	&	$	80.7	_{-	2.2	}	^{+	2.4	}	$	&	$	1.91	_{-	0.13	}	^{+	0.15	}	$	&	$	50.0	\pm	9.6	$	\\
39	&	$	0.85	\pm	0.01					$	&	$	78.6	_{-	2.2	}	^{+	1.5	}	$	&	$	2.01	_{-	0.13	}	^{+	0.11	}	$	&	$	88.3	\pm	29.7	$	\\
40	&	$	0.83	_{-	0.01	}	^{+	0.02	}	$	&	$	66.8	_{-	4.8	}	^{+	2.2	}	$	&	$	2.13	\pm	0.09					$	&	$	-			$	\\
41	&	$	0.84	_{-	0.02	}	^{+	0.01	}	$	&	$	61.2	_{-	2.3	}	^{+	4.3	}	$	&	$	2.40	_{-	0.03	}	^{+	0.06	}	$	&	$	-			$	\\
42	&	$	0.81	_{-	0.02	}	^{+	0.01	}	$	&	$	64.2	_{-	3.6	}	^{+	1.9	}	$	&	$	2.43	_{-	0.03	}	^{+	0.02	}	$	&	$	-			$	\\
43	&	$	0.85	_{-	0.01	}	^{+	0.02	}	$	&	$	72.0	_{-	10.4	}	^{+	1.2	}	$	&	$	2.10	_{-	0.13	}	^{+	0.06	}	$	&	$	117.4	\pm	41.4	$	\\
44	&	$	0.93	\pm	0.01					$	&	$	54.4	_{-	3.7	}	^{+	2.3	}	$	&	$	2.21	\pm	0.10					$	&	$	86.5	\pm	18.3	$	\\
45	&	$	0.93	\pm	0.01					$	&	$	54.4	_{-	3.7	}	^{+	2.3	}	$	&	$	2.21	\pm	0.10					$	&	$	86.5	\pm	18.3	$	\\
46	&	$	0.85	\pm	0.03					$	&	$	44.7	_{-	4.5	}	^{+	5.2	}	$	&	$	2.37	_{-	0.06	}	^{+	0.04	}	$	&	$	162.9	\pm	45.2	$	\\
47	&	$	0.86	_{-	0.04	}	^{+	0.02	}	$	&	$	47.2	_{-	2.3	}	^{+	2.4	}	$	&	$	2.42	_{-	0.05	}	^{+	0.08	}	$	&	$	135.7	\pm	28.0	$	\\
																																					
\end{tabular}																																					
\end{center}																																					
\end{table*}																																					
																																					
\begin{table*}																																					
\renewcommand{\arraystretch}{1.3}																																					
\begin{center}																																					
\begin{tabular}{|c|c|c|c|c|}																																					
\hline																																					
$\#$	&	kT	(keV)								&		R	(Km)							&			$\Gamma$							&		E$_{c}$				\\
\hline																																					
\hline																																					
48	&	$	0.86	\pm	0.03					$	&	$	41.6	_{-	3.7	}	^{+	5.7	}	$	&	$	2.37	_{-	0.04	}	^{+	0.05	}	$	&	$	200.8	\pm	46.8	$	\\
49	&	$	0.84	_{-	0.04	}	^{+	0.03	}	$	&	$	43.5	_{-	3.6	}	^{+	7.3	}	$	&	$	2.28	_{-	0.03	}	^{+	0.07	}	$	&	$	79.0	\pm	9.8	$	\\
50	&	$	0.84	_{-	0.02	}	^{+	0.03	}	$	&	$	50.1	_{-	4.4	}	^{+	3.4	}	$	&	$	2.43	_{-	0.07	}	^{+	0.06	}	$	&	$	99.1	\pm	28.5	$	\\
51	&	$	0.85	_{-	0.04	}	^{+	0.01	}	$	&	$	47.2	_{-	2.2	}	^{+	4.3	}	$	&	$	2.60	_{-	0.01	}	^{+	0.05	}	$	&		-				\\
52	&	$	0.78	_{-	0.02	}	^{+	0.01	}	$	&	$	61.6	_{-	4.2	}	^{+	5.0	}	$	&	$	2.43	_{-	0.05	}	^{+	0.07	}	$	&		-				\\
53	&	$	0.78	\pm	0.02					$	&	$	60.2	_{-	4.2	}	^{+	5.7	}	$	&	$	2.43	_{-	0.11	}	^{+	0.07	}	$	&		-				\\
54	&	$	0.80	\pm	0.02					$	&	$	59.0	\pm	4.2					$	&	$	2.17	_{-	0.07	}	^{+	0.06	}	$	&		-				\\
55	&	$	0.77	\pm	0.02					$	&	$	63.6	_{-	4.2	}	^{+	5.0	}	$	&	$	2.44	_{-	0.04	}	^{+	0.08	}	$	&		-				\\
56	&	$	0.77	_{-	0.01	}	^{+	0.02	}	$	&	$	65.2	_{-	4.0	}	^{+	2.5	}	$	&	$	2.42	_{-	0.10	}	^{+	0.07	}	$	&		-				\\
57	&	$	0.77	_{-	0.02	}	^{+	0.01	}	$	&	$	64.0	_{-	2.9	}	^{+	7.3	}	$	&	$	2.34	_{-	0.04	}	^{+	0.10	}	$	&		-				\\
58	&	$	0.74	_{-	0.01	}	^{+	0.02	}	$	&	$	70.5	_{-	4.9	}	^{+	3.9	}	$	&	$	2.40	_{-	0.03	}	^{+	0.06	}	$	&		-				\\
59	&	$	0.75	_{-	0.03	}	^{+	0.00	}	$	&	$	66.7	_{-	2.0	}	^{+	5.9	}	$	&	$	2.45	_{-	0.07	}	^{+	0.10	}	$	&		-				\\
60	&	$	0.75	_{-	0.02	}	^{+	0.01	}	$	&	$	65.5	_{-	3.3	}	^{+	4.1	}	$	&	$	2.32	\pm	0.06					$	&		-				\\
61	&	$	0.75	_{-	0.01	}	^{+	0.02	}	$	&	$	66.2	_{-	4.1	}	^{+	4.8	}	$	&	$	2.28	\pm	0.08					$	&		-				\\
62	&	$	0.76	\pm	0.02					$	&	$	61.5	_{-	4.5	}	^{+	4.7	}	$	&	$	2.26	\pm	0.09					$	&		-				\\
63	&	$	0.76	\pm	0.02					$	&	$	57.7	_{-	4.8	}	^{+	2.5	}	$	&	$	2.36	\pm	0.05					$	&		-				\\
64	&	$	0.75	_{-	0.02	}	^{+	0.01	}	$	&	$	59.5	_{-	3.6	}	^{+	5.2	}	$	&	$	2.41	_{-	0.05	}	^{+	0.03	}	$	&		-				\\
65	&	$	0.73	_{-	0.01	}	^{+	0.02	}	$	&	$	66.8	_{-	5.0	}	^{+	3.0	}	$	&	$	2.39	_{-	0.05	}	^{+	0.03	}	$	&		-				\\
66	&	$	0.80	_{-	0.03	}	^{+	0.01	}	$	&	$	47.2	_{-	2.2	}	^{+	2.8	}	$	&	$	2.42	_{-	0.02	}	^{+	0.05	}	$	&		-				\\
67	&	$	0.80	_{-	0.02	}	^{+	0.01	}	$	&	$	46.0	_{-	2.8	}	^{+	1.6	}	$	&	$	2.47	_{-	0.02	}	^{+	0.04	}	$	&		-				\\
68	&	$	0.81	\pm	0.02					$	&	$	43.5	_{-	3.1	}	^{+	4.6	}	$	&	$	2.45	_{-	0.01	}	^{+	0.03	}	$	&		-				\\
69	&	$	0.78	\pm	0.02					$	&	$	49.5	_{-	3.5	}	^{+	4.5	}	$	&	$	2.48	_{-	0.04	}	^{+	0.07	}	$	&		-				\\
70	&	$	0.77	_{-	0.01	}	^{+	0.03	}	$	&	$	49.7	_{-	5.0	}	^{+	2.9	}	$	&	$	2.46	\pm	0.04					$	&		-				\\
71	&	$	0.73	_{-	0.01	}	^{+	0.02	}	$	&	$	58.0	_{-	5.1	}	^{+	4.0	}	$	&	$	2.35	\pm	0.04					$	&		-				\\
72	&	$	0.79	\pm	0.02					$	&	$	43.5	_{-	2.9	}	^{+	3.5	}	$	&	$	2.41	_{-	0.02	}	^{+	0.03	}	$	&		-				\\
73	&	$	0.77	_{-	0.02	}	^{+	0.01	}	$	&	$	45.1	_{-	2.6	}	^{+	3.2	}	$	&	$	2.40	_{-	0.02	}	^{+	0.04	}	$	&		-				\\
74	&	$	0.73	_{-	0.02	}	^{+	0.01	}	$	&	$	49.7	_{-	3.1	}	^{+	4.6	}	$	&	$	2.28	\pm	0.04					$	&		-				\\
75	&	$	0.69	_{-	0.01	}	^{+	0.02	}	$	&	$	62.5	_{-	4.7	}	^{+	2.7	}	$	&	$	2.18	\pm	0.07					$	&		-				\\
76	&	$	0.70	_{-	0.01	}	^{+	0.02	}	$	&	$	57.8	_{-	4.9	}	^{+	2.8	}	$	&	$	2.09	_{-	0.08	}	^{+	0.05	}	$	&		-				\\
77	&	$	0.67	_{-	0.03	}	^{+	0.01	}	$	&	$	61.0	_{-	13.5	}	^{+	5.4	}	$	&	$	2.00	_{-	0.09	}	^{+	0.08	}	$	&		-				\\
78	&	$	0.68	_{-	0.01	}	^{+	0.02	}	$	&	$	52.5	_{-	5.1	}	^{+	4.1	}	$	&	$	2.13	_{-	0.07	}	^{+	0.21	}	$	&		-				\\
79	&	$	0.68	_{-	0.01	}	^{+	0.02	}	$	&	$	45.6	_{-	4.5	}	^{+	1.8	}	$	&	$	2.17	_{-	0.11	}	^{+	0.06	}	$	&		-				\\
80	&	$	0.67	_{-	0.01	}	^{+	0.02	}	$	&	$	45.3	_{-	4.0	}	^{+	4.2	}	$	&	$	2.10	_{-	0.03	}	^{+	0.05	}	$	&		-				\\
81	&	$	0.87	_{-	0.04	}	^{+	0.06	}	$	&	$	20.0	_{-	1.8	}	^{+	4.0	}	$	&	$	2.12	_{-	0.13	}	^{+	0.09	}	$	&		-				\\
82	&	$	0.73	_{-	0.02	}	^{+	0.03	}	$	&	$	32.9	_{-	4.8	}	^{+	1.7	}	$	&	$	2.23	_{-	0.05	}	^{+	0.06	}	$	&		-				\\
83	&	$	0.70	_{-	0.02	}	^{+	0.03	}	$	&	$	34.4	_{-	4.1	}	^{+	3.5	}	$	&	$	2.13	\pm	0.05					$	&		-				\\

\hline																																					
\end{tabular}																																					
\caption{Spectral parameters. Columns are: observation number, inner disk temperature, inner disk radius (assuming a distance of 8 kpc and an inclination of 60$^\circ$, and high-energy cutoff.}\label{tab:spettrali}																																					
\end{center}																																					
\end{table*}

\begin{table*}
\renewcommand{\arraystretch}{1.3}
\begin{center}
\begin{tabular}{|c|c|c|c|c|}
\hline
			&    	H	&	SPL 	&      TD	& INT\\
\hline
\hline

LHS			&      17 	&	-	&	-	&	2	\\
HIMS		&	4	&	3	&	-	&	7	\\
SIMS		&	-	&	3	&	-	&	1	\\
HSS			&	-	&	2	&	19	&	25	\\

\hline
\end{tabular}
\caption{Comparison between the two classificatin schemes of Belloni (2009) and McRem (see text). In the first row we report the states of the McClintock \& Remillard classification: Hard (H), Steep Power Law (SPL), Thermal Dominated (TD) and Intermediate (INT); in the first column we report the states of the classification we used: Low Hard State (LHS), Hard Intermediate State (HIMS), Soft Intermediate State (SIMS), High Soft State (HSS). We see that more than 40\% of the observations results unclassified, that is belong to the Intermediate state. The states coming from the two different classifications are also reported in Tab. \ref{tab:colori}.}\label{tab:remillard}
\end{center}
\end{table*}


\begin{thebibliography}{}


\bibitem[]{} Belloni T., 2005 AIPC, 797, 197

\bibitem[]{} Belloni T., Hasinger G., 1990, A\&A, 230, 103

\bibitem[]{} Belloni T., van der Klis M., Lewin W.H.G., van Paradijs J., Dotani T., et al., 1997, A\&A, 322, 857

\bibitem[]{} Belloni T., M\'endez M., van der Klis M., Lewin W.H.G., Dieters S., 1999, ApJ, 519, L159

\bibitem[]{} Bellon T., Homan J., Cui W., Swank J., 2004, ATel \#236

\bibitem[]{} Belloni T., Homan J., Casella P., van der Klis M., Nespoli E., et al., 2005, A\&A, 440, 207

\bibitem[]{} Belloni T., Parolin I., Del Santo M., Homan J., Casella P., et al., 2006, MNRAS, 367, 1113

\bibitem[]{} Brocksopp C., Fender R. P., McCollough M., Pooley G. G., Rupen M. P. et al., 2002 MNRAS, 331, 765

\bibitem[]{} Buxton M., Bailyn C., 2004a, ATel, 270

\bibitem[]{} Caballero-Garc\'ia M. D., Miller J. M., Trigo M., Kuulkers E., Fabian A. C., et al. 2009, ApJ, 692, 1339

\bibitem[]{} Cadolle Bel M., Rodriguez J., Goldwurm A., Goldoni P., Laurent P., et al., 2005, ATel, 574, 1

\bibitem[]{} Casella P., Belloni T., Homan J., Stella L., 2004 cosp, 35, 3875

\bibitem[]{} Casella P., Belloni T., Stella L., 2005 ApJ, 629, 403

\bibitem[]{} Corbel S., Nowak M.A., Fender R.P., Tzioumis A.K., Markoff S., 2003, A\&A, 400, 1007

\bibitem[]{} Corongiu A., Chiappetti L., Haardt F., Treves A., Colpi M., Belloni T., 2003, A\&A, 408, 347

\bibitem[]{} Del Santo M., Bazzano A., Zdziarski A. A., et al., 2005, A\&A, 433, 613

\bibitem[]{} Del Santo M., Belloni T. M., Homan J., Bazzano A., Casella P. et al.,  2009, MNRAS, 392, 992

\bibitem[]{} Del Santo M., Malzac J., Jourdain E., Belloni T., Ubertini P., 2008, MNRAS, 390, 227

\bibitem[]{} Dove J.B., Wilms J., Maisack M., Begelman M.C., 1997,  ApJ, 487, 759 

\bibitem[]{} Dove J.B., Wilms J., Nowak M.A., Vaughan B.A., Begelman M.C., 1998, MNRAS, 298, 729

\bibitem[]{} Fender R.P., Corbel S., Tzioumis T., McIntyre V., Campbell-Wilson D. et al., 1999, ApJ, 519, L165

\bibitem[]{} Fender R.P., Corbel S., Tzioumis T., Tingay S., Brocksopp C., Gallo E., 2002, ATel, 107

\bibitem[]{} Fender R.P. Belloni T., Gallo E., 2004, MNRAS, 355, 1105

\bibitem[]{} 2009, Fender, R. P., Homan, J., Belloni, T. M., MNRAS, 396, 1370

\bibitem[]{} Fender R.P., Corbel S., Tzioumis T., Tingay S., Brocksopp C., Gallo E., 2002, ATel, \#107

\bibitem[]{} Frontera F., Palazzi E., Zdziarski A.A., Haardt F., Perola G.C. et al.,  2001, ApJ, 546, 1027

\bibitem[]{} Gallo E., Corbel S., Fender R.P., Maccarone T.J., Tzioumis A.K., 2004, MNRAS, 347, L52

\bibitem[]{} George I., Fabian A.C., 1991, MNRAS, 249, 352

\bibitem[]{} Giannios D., 2005, A\&A, 437, 1007

\bibitem[]{} Gierlinski \& Newton, 2006, MNRAS, 370, 837

\bibitem[]{} Gierliski M., Zdziarski A. A., 2005 MNRAS, 363, 1349

\bibitem[]{} Gierlinski M., Zdziarski A.A., Poutanen J., Coppi P.S., Ebisawa K., Johnson W.N., 1999, MNRAS, 309, 496

\bibitem[]{} Grove J.E., Johnson W.N., Kroeger R.A., McNaron Brown K., Skibo J.G. et al., 1998, ApJ, 500, 899

\bibitem[]{} Hannikainen D.C., Hunstead R.W., Campbell-Wilson D., Sood R.K., 1998, A\&A, 337, 460

\bibitem[]{} Hynes R. I., Steeghs D., Casares J., Charles P. A., OÕBrien K., 2003, ApJ, 583, L95

\bibitem[]{} Hynes R. I., Steeghs D., Casares J., Charles P. A., OÕBrien K., 2004, ApJ, 609, 317

\bibitem[]{} Homan J., Wijnands R., 2002, ATel, 109, 1

\bibitem[]{} Homan J., Buxton M., Markoff S., Bailyn C. D., Nespoli E. et al. ,  2005, ApJ, 624, 295

\bibitem[]{} Homan J., Buxton M., Markoff S., Bailyn C., Nespoli E. et al., 2004, HEAD, 8, 1511

\bibitem[]{} Ilovaisky S.A., Chevalier C., Motch C., Chiappetti L., 1986, A\&A, 164, 67

\bibitem[]{} Israel G., Covino S., Kuulkers E., Zerbi F.M., Chincarini G., Rodon\'o M., Antonelli L.A., 2004, ATel, 243

\bibitem[]{} Joinet A., Kalemci E., Senziani F., 2008, ApJ, 679, 655

\bibitem[]{} Kong A.K.H., Kuulkers E., Charles P.A., Smale A.P., 2000, MNRAS, 311, 405S

\bibitem[]{} Kuulkers E., Bodaghee A., Foschini L., Guainazzi M.,Matt, G., Israel G., Nicastro F., et al., 2004, ATel, 240

\bibitem[]{} Krimm H. A., Barbier L., Barthelmy S. D., Cummings J., Fenimore E. et al., 2006, ATel, 968, 1

\bibitem[]{} Laurent P. Titarchuk L., 1999, ApJ, 511, 289

\bibitem[]{} Leahy D.A., Darbro W., Elsner R.F., Weisskopf M.C., Kahn S., Sutherland P.,G., Grindlay J.E., 1983, ApJ, 266, 160 

\bibitem[]{} Maejima Y., Makishima K., Matsuoka M., Ogawara Y., Oda M., Tawara Y., Doi, K., 1984, ApJ, 285, 712

\bibitem[]{} Malzac J., Petrucci P.O., Jourdain E., Cadolle Bel M., Sizun P. et al., 2005, A\&A, Submitted

\bibitem[]{} Markoff S., Falcke H., Fender R.P., 2001, A\&A, 372, L25 

\bibitem[]{} Markoff S., Nowak M.A., Corbel S., Fender R.P., Falcke H., 2003, A\&A, 397, 645

\bibitem[]{} Markoff S., Nowak M.A., Wilms J., 2005, ApJ, 635,1203

\bibitem[]{}  McClintock J.E., Remillard R.A., Rupen M.P., Torres M.A.P., Steeghs D. et al., 2009, arXiv:0705.1034v3

\bibitem[]{} M\'endez M., van der Klis M., 1997, ApJ, 479, 926

\bibitem[]{} Miller J.M., Fabian A.C., Wijnands W., Remillard R.A., Wojodowski P. et al., 2002, ApJ, 578, 348 

\bibitem[]{} Miller J.M., Fabian A.C., Reynolds C.S, Nowak M.A., Homan J. et al.,  2004a, ApJ, 606, L131 

\bibitem[]{} Miller J. M., Raymond J., Fabian A. C., Homan J., Nowak M.A. et al.,  2004b, ApJ, 601, 450 

\bibitem[]{} Miller et. al, 2006, Nature, 441,953)

\bibitem[]{} Miller et al, 2008, APJ, 679, L113

\bibitem[]{} Miyakawa T., Yamaoka K., Homan J., Saito K., Dotani T. et al., 2008, PASJ, 60, 637

\bibitem[]{} Miyamoto S., Kimura K., Kitamoto S., Dotani T., Ebisawa K., 1991, ApJ, 383, 784S., 2004, MNRAS, 351, 791 

\bibitem[]{} Nespoli E., Belloni T., Homan J., Miller J. M., Lewin W. H. G. et al., 2003A\&A, 412, 235

\bibitem[]{} Neilsen \& Lee, 2009, Nature, 458, 481

\bibitem[]{} Nowak M.A., Wilms J., Dove J.B., 1999, ApJ, 517, 355

\bibitem[]{} Nowak M.A., Wilms J., Dove J.B., 2002, MNRAS, 332, 856

\bibitem[]{}  McClintock J.E., Remillard R.A., Cambridge University Press, 2006, pp 157-214

\bibitem[]{} Remillard R.A. \& McClintok J.E., 2006, ARA\&A, 44, 49

\bibitem[]{} Reynolds C. S., Nowak, M. A., 2003, PhR, 377, 389

\bibitem[]{} Swank J. H., Smith E. A., Smith D. M., Markwardt C. B., 2006, ATel, 944, 1

\bibitem[]{} Smith D. M., Swank J. H., Heindl W. A., Remillard R. A., 2002a, ATel, 85

\bibitem[]{} Smith D. M., Belloni T., Heindl W. A., et al., 2002c, ATel, 95

\bibitem[]{} Smith D. M., Heindl W. A., Swank J. H., Wilms J., Pottschmidt K., 2004a, ATel, 231

\bibitem[]{} Sunyaev R. A.,  Titarchuk L. G. ,1980 A\&A, 86, 121

\bibitem[]{} Sunyaev R. A., Truemper J., 1979, Nature, 279, 506

\bibitem[]{} Titarchuk L.,  1994 ApJ, 434, 570

\bibitem[]{} Turolla R., Zane S., Titarchuk L., 2002, ApJ, 576, 349

\bibitem[]{} Wilms J., Nowak M.A., Dove J.B., Fender R.P., Di Matteo T., 1999, ApJ, 522, 460

\bibitem[]{} Wilms J., Nowak M.A., Pottschmidt K., Pooley G.G., Fritz S., 2006, A\&A, 447, 245

\bibitem[]{} Zhang W., Jahoda K., Swank J.H., Morgan E.H., Giles A.B., 1995, ApJ, 449, 930

\bibitem[]{} Zdziarski et al., 1996, A\&A Suppl. 120, 553

\bibitem[]{} Zdziarski A.A., Grove J.E., Poutanen J., Rao A.R.,  Vadawale V.,  2001, ApJ, 554, L45

\bibitem[]{} Zdziarski A.A., Gierlinski M., Mikola jewska J., Wardzinski G., Smith D.M. et al., 2004, MNRAS, 351, 791 


\end{thebibliography}
\end{document}